%% file: main.tex
\documentclass[journal=jacsat,manuscript=article]{achemso}

\usepackage[version=3]{mhchem} 
\usepackage{xr}
\makeatletter
\newcommand*{\addFileDependency}[1]{
  \typeout{(#1)}
  \@addtofilelist{#1}
  \IfFileExists{#1}{}{\typeout{No file #1.}}
}
\makeatother

\newcommand*{\myexternaldocument}[1]{%
    \externaldocument{#1}%
    \addFileDependency{#1.tex}%
    \addFileDependency{#1.aux}%
}
\myexternaldocument{supporting_info}
\usepackage{hyperref}
\hypersetup{
    colorlinks=true,
    linkcolor=blue,
    citecolor=blue,
    filecolor=magenta,      
    urlcolor=blue,
    pdfpagemode=FullScreen,
    }

\newcommand{\bfv}[1]{{\mbox{\boldmath{$#1$}}}}
\newcommand{\x}{\bfv{x}}

\author{Wei-Tse Hsu}
\affiliation{Department of Chemical and Biological Engineering, University of Colorado Boulder, Boulder, CO 80305}
\author{Valerio Piomponi}
\affiliation{Scuola Internazionale Superiore di Studi Avanzati, via Bonomea 265, 34136 Trieste, Italy}
\author{Pascal T. Merz}
\affiliation{Department of Chemical and Biological Engineering, University of Colorado Boulder, Boulder, CO 80305}
\author{Giovanni Bussi}
\affiliation{Scuola Internazionale Superiore di Studi Avanzati, via Bonomea 265, 34136 Trieste, Italy}
\author{Michael R. Shirts}
\affiliation{Department of Chemical and Biological Engineering, University of Colorado Boulder, Boulder, CO 80305}
\email{michael.shirts@colorado.edu}

\title
  {Adding alchemical variables to metadynamics to enhance sampling in free energy calculations}


\begin{document}



\begin{abstract}
Performing alchemical transformations, in which one molecular system is nonphysically changed to another system, is a popular approach adopted in performing free energy calculations associated with various biophysical processes, such as protein-ligand binding or the transfer of a molecule between environments. While the sampling of alchemical intermediate states in either parallel (e.g. Hamiltonian replica exchange) or serial manner (e.g. expanded ensemble) can bridge the high-probability regions in the configurational space between two end states of interest, alchemical methods can fail in scenarios where the most important slow degrees of freedom in the configurational space are in large part orthogonal to the alchemical variable, or if the system gets trapped in a deep basin extending in both the configurational and alchemical space. 

To alleviate these issues, we propose to use alchemical variables as an additional dimension in metadynamics, augmenting the ability both to sample collective variables and to enhance sampling in free energy calculations. In this study, we validate our implementation of ``alchemical metadynamics'' in PLUMED with test systems and alchemical processes with varying complexities and dimensions of collective variable space, including the interconversion between the torsional metastable states of a toy system and the methylation of a nucleoside both in the isolated form and in a duplex. We show that multi-dimensional alchemical metadynamics can address the challenges mentioned above and further accelerate sampling by introducing configurational collective variables. The method can trivially be combined with other metadynamics-based algorithms implemented in PLUMED. The necessary PLUMED code changes have already been released for general use in PLUMED 2.8.
\end{abstract}

\section{Introduction}
With the fast advent of high-performance computing and parallel computing over the years, methods such as molecular dynamics (MD) and Monte Carlo (MC) simulations have become increasingly useful in elucidating the dynamics of transformation processes of condensed matter systems. They are most useful when the system can sample efficiently from all the energetically relevant conformations, in which case we can extract valuable thermodynamic and structural information about the system, such as the solvation free energy of a molecule or the binding ensemble of a binding complex. However, comprehensive sampling in the phase space is generally challenging in traditional MD simulations because the system must rely on very rare fluctuations to cross the free energy barriers that separate metastable states of interest. In most systems of interest, this low transition probability between metastable states makes the timescale required for achieving system ergodicity in unbiased sampling impractically long. 

To address this challenge, a wide variety of advanced sampling methods have been developed in the past decades~\cite{henin2022enhanced}. One particular flavor of advanced sampling methods involves sampling along a set of predefined coarse-grained descriptors of the system, or collective variables (CVs). Traditionally, CVs could be any function of the atomic coordinates of the system, but the optimal ones should correspond to the slowest degrees of freedom that distinguish different metastable states. Methods relying on the use of CVs include umbrella sampling~\cite{umbrella}, adaptive biasing force~\cite{ABF}, metadynamics~\cite{metad} and their variations~\cite{bussi2006free, bonomi2010enhanced, var2}. 

Another category of advanced sampling methods is known as generalized ensemble algorithms, which includes simulated tempering~\cite{marinari1992simulated}, replica exchange~\cite{TREMD, HREMD}, expanded ensemble~\cite{EXE}, $\lambda$-dynamics~\cite{knight2009lambda}, and their variations~\cite{oshima2019replica, knight2011multisite}. These methods do not require any predefined CVs, but a series of intermediate or auxiliary states with modified probabilities of the coordinates of the system. These states are typically defined by varying the temperature or the Hamiltonian of the system. The motivation to introduce these states is often physical; for example, alchemically connecting two end states with a molecule deleted or ``mutated'' is usually the most efficient way to calculate many free energy differences~\cite{mey2020best}. For example, the sampling in the temperature space allows us to determine thermodynamic observables of interest as a function of temperature, while simulations with alchemical intermediate states are useful for calculating the free energy difference between the end states of alchemical processes. In free energy calculations, sampling in alchemical intermediate states obviates the need of defining CVs, which could be non-trivial in systems where the slowest-relaxing coordinates are not intuitive, such as the escape of a ligand from a buried binding pocket~\cite{buried1, buried2}. 

\begin{figure}[ht]
    \centering
    \includegraphics[width=\textwidth]{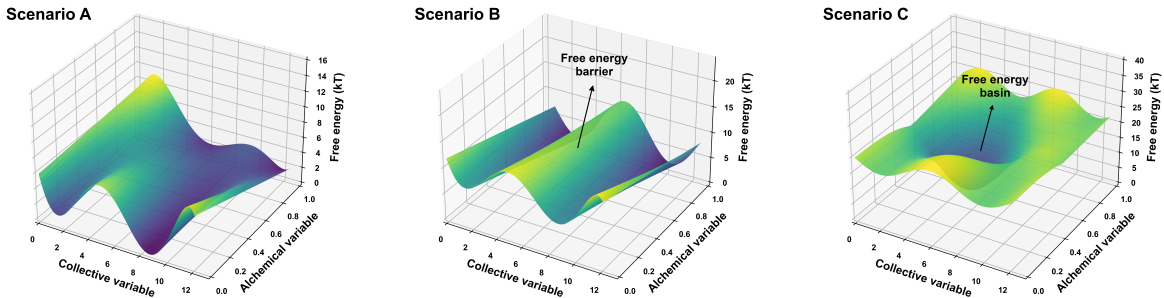}   
    \caption{Common scenarios of sampling along the alchemical direction in the phase space. In scenario A, the free energy barrier present at $\lambda=0$ is absent from larger $\lambda$ values, so the system can go around the free energy barrier by pure alchemical sampling. On the other hand, pure alchemical bias would fail to accelerate configurational sampling in scenarios B and C. In scenario B, the free energy barrier extending across all $\lambda$ states prevents the system from sampling both metastable states at $\lambda$ being 0. In scenario C, the free energy basin in both the alchemical and configurational directions can trap the system during the adaptive build-up of the alchemical weights.}
    \label{FES_scenarios}
\end{figure}

However, these additional states can also remove or lessen the kinetic barriers with states of interest at the intermediate states, either by the intentional choice of additional states, or as a useful side effect. As the system jointly samples the coordinate/configurational space and this additional sampling direction, nonphysical pathways are created allowing the system to circumvent free energy barriers in the configurational space (scenario A in Figure \ref{FES_scenarios}). The increased probability overlap between adjacent intermediate states can often enhance the diffusion in not only the temperature/Hamiltonian direction, but also the configurational space. In replica exchange, these states are sampled with the ensembles progressing forwards in time in parallel, while in serial approaches, a single simulation can move between states in either discrete (expanded ensemble) or continuous space ($\lambda$-dynamics). In order to have even sampling between the states, one must add biases or weights to the higher free energy states so they can be sampled. These weights, which are absent in replica exchange, similarly modify the underlying free energy surface as the biasing potentials in metadynamics do. 

Although sampling in the alchemical variable can create new ensembles where the slowest physical collective variables are no longer so slow, it cannot necessarily enhance sampling where the configurational barriers are almost orthogonal to the alchemical direction. For example, a configurational free energy barrier can be present for all the alchemical states (scenario B in Figure \ref{FES_scenarios}) so that the system could remain stuck even if it is able to drift to other alchemical states or cross free energy barriers along the alchemical direction. Another scenario that could possibly trap the system is the presence of large free energy basins in both the configurational and alchemical directions (scenario C in Figure \ref{FES_scenarios}). Importantly, upon the application of the alchemical bias, even if the free energy landscape in the alchemical direction can be flat for a range of configurational CV values, a flat sampling distribution in coordinate space is not guaranteed, i.e. the sampling in the configurational space can remain limited.

Currently, the challenges mentioned above can be addressed to varying extents. For example, the alchemical flying Gaussian method~\cite{trapl2020prediction} or the conveyor belt method combined with specific biasing~\cite{hahn2020overcoming} do not bias the alchemical but only configurational space in alchemical sampling. Although these methods might have difficulties in getting the system out of free energy basins in scenario C in Figure \ref{FES_scenarios}, they might be sufficient to overcome the extended configurational free energy barrier in scenario B in Figure \ref{FES_scenarios}. Methods such as $\lambda$ local elevation umbrella sampling ($\lambda$-LEUS)~\cite{bieler2014local, bieler2015multistate}, orthogonal space random walk (OSRW)~\cite{zheng2008random}, double-integration orthogonal space tempering (DI-OST)~\cite{zheng2012practically}, and adaptive landscape flattening (ALF)~\cite{hayes2017adaptive}, which all work with continuous alchemical space in their proposals, are able to apply biasing potentials in both the alchemical or configurational directions. Theoretically, they can facilitate the escape of the system from the deep free energy basins shown in scenario B in Figure \ref{FES_scenarios}, but each of them was implemented in a different context. In addition, their algorithmic designs and allowed forms of configurational bias tend to be specific rather than general. For example, OSRW does not generalize well to multiple dimensions because finding a CV that is simultaneously orthogonal to $dH/d\lambda$ and $\lambda$ is generally difficult.

In light of the need for a more generalized approach that can address the issues in scenarios B and C in Figure \ref{FES_scenarios}, especially in cases where multi-dimensional biases in the configurational space are needed, we propose to use alchemical variables as an additional dimension in metadynamics. Although an alchemical variable is not a collective variable of coordinates whose values divide the coordinate space into disjoint sets like a center of mass distance or radius of gyration, it has an associated free energy and dynamics, and thus can fit into the same formalism.  We term this approach \textit{alchemical metadynamics} and have implemented it in PLUMED 2.8~\cite{tribello2014plumed} (initially, only for GROMACS). Given the well-developed library and flexible syntax that PLUMED has for defining configurational CVs and various types of restraints, the implementation of alchemical metadynamics within PLUMED is particularly useful for smoothly flattening highly-dimensional free energy landscapes in a more general way compared to methods such as $\lambda$-LEUS, OSRW, or DI-OST. To demonstrate the usage of such an algorithm and validate its implementation in PLUMED, we employed this method to estimate the free energy difference of different alchemical processes as elaborated in the Methods section, from decoupling an argon atom or a molecule composed 4 interaction sites, to the methylation of a nucleobase for both  the isolated form and in a duplex. 

\section{Theory}
\subsection{Metadynamics}
As one of the most popular CV-based advanced sampling methods, metadynamics~\cite{metad} accelerates the sampling by depositing Gaussian biasing potentials to the underlying free energy surface of the system. The biasing potential is a function of the vector of collective variables of interest $\bfv{\xi}$, which can be regarded as a reduced dimensional space calculated from a configuration $\x$ by the mapping $\Phi(\x)=\bfv{\xi}$. During the simulation, biasing potentials are deposited to seek roughly equal sampling across the reduced dimensional space of interest. Let the CV vector $\bfv{\xi}$ be $d$-dimensional, i.e. $\bfv{\xi}=(\xi_{1}(\x), \xi_{2}(\x), ..., \xi_{d}(\x))$. The total biasing potential added after a period of time $t$ can be expressed as 

\begin{equation}
    V(\bfv{\xi}, t) = W \sum_{t'=k\tau, k \in N}^{t'<t}\exp \left( -\sum_{i=1}^{d}\frac{\left(\xi_{i}-\xi_{i}(\x (k\tau))\right)^{2}}{2\sigma_{i}^{2}}\right)
\end{equation}
where $W$ is the height of the Gaussian, $k$ is the number of Gaussian depositions, $\tau$ is the deposition stride and $\sigma_{i}$ is the width of the Gaussian along the $i$-th dimension. Notably, the Gaussian height $W$ can be either constant (in standard metadynamics) or time-dependent (in well-tempered metadynamics~\cite{WTMetaD}) during the course of the simulation, with the latter more commonly adopted for a smoother convergence and better concentration on the physically relevant regions of the configurational space. Specifically, the time-dependent Gaussian height $W(k\tau)$ can be written as:

\begin{equation}
    W(k\tau)=W_{0}\exp \left(-\frac{V(\vec{\xi}(\x(k\tau)), k\tau)}{k_{B}\Delta T} \right)
\end{equation}
where $W_0$ is the initial Gaussian height and $\Delta T$ is a temperature parameter that incorporates a user-defined bias factor $\gamma=(T + \Delta T)/T$ for adjusting the decay rate of the bias. In well-tempered metadynamics, the free energy surface as a function of the multi-dimensional CV $\bfv{\xi}$ can be estimated by the following relationship:

\begin{equation}
    V(\bfv{\xi}, t \rightarrow \infty)=-\frac{\Delta T}{T + \Delta T}F(\bfv{\xi})=-(1-\frac{1}{\gamma})F(\bfv{\xi})
\end{equation}

To efficiently obtain a reasonable estimate of the free energy difference of interest, the dimensions of the chosen set of CVs must be as low as possible while still capturing the slowest degrees of freedom of the system, as the space to be explored increases exponentially with the number of CVs, leading to prohibitive time to converge weights. For multi-dimensional metadynamics, introducing multiple interacting walkers~\cite{multi-metaD} to sample the same free energy surface along different dimensions of the CVs can be a useful strategy for speeding up the reconstruction and exploration of the free energy surface.  

\subsection{Alchemical metadynamics}
In alchemical metadynamics, the alchemical variable $\lambda$ is introduced in the generalized CV vector $\bfv{\xi'}=(\lambda, \xi_{1}(\x), \xi_{2}(\x), ..., \xi_{d}(\x))$ such that the joint space of $\lambda$ and $\bfv{\xi}$ is sampled with the aid of the biasing potential $V(\bfv{\xi'})$. Unlike the configurationally defined CVs, the alchemical variable is not a function of atomic coordinates. In the current implementation, we assume that the alchemical variable takes discrete values, i.e. the state index that can be mapped to a vector of coupling parameters for decoupling/switching different interactions, such as van der Waals interactions, electrostatic interactions, or any kind of restraints. Similarly to expanded ensemble, alchemical metadynamics alternates the sampling along the alchemical direction and the coordinate directions. The sampling in the discrete alchemical space can be done by Monte Carlo sampling just as the alchemical sampling in expanded ensemble, while the coordinate direction is sampled by molecular dynamics as in any other type of metadynamics. Currently, our implementation of alchemical metadynamics is available in the combination of PLUMED 2.8 interfaced with GROMACS 2020, and in any combination of more recent versions of each. When using alchemical metadynamics, the state index of the alchemical or coupling parameter $\lambda$ is passed from GROMACS to PLUMED along with the system configuration required to compute configurational CVs. PLUMED uses these alchemical indices and any other CV present to track the visited states of the system and calculate the metadynamics bias, while GROMACS updates the alchemical state via MC. When calculating the energy of the current and the candidate $\lambda$ states, GROMACS includes the metadynamics bias provided by PLUMED. This approach is compatible with all MC schemes in alchemical space offered by GROMACS, including the Metropolis-Hastings algorithm~\cite{hastings1970monte}, Barker transition method~\cite{barker1965monte}, Gibbs sampling~\cite{geman1984stochastic, liu2001monte}, and Metropolized-Gibbs sampling~\cite{liu1996peskun, chodera2016simple}.

Theoretically, one-dimensional alchemical metadynamics, which does not apply configurational but only alchemical bias, is effectively equivalent to expanded ensemble with a different weight updating procedure for allowing roughly equal sampling across alchemical states. For example, in an expanded ensemble where the Wang-Landau algorithm~\cite{desgranges2012evaluation, wang2001efficient, belardinelli2007fast} is used for weight updating, the reduced potential of the system is incremented by a Wang-Landau incrementor whenever a move across alchemical states is attempted. This is analogous to the periodic deposition of Gaussian potentials in 1D alchemical metadynamics, especially when the Gaussian deposition stride is the same as the number of integration steps between attempted moves in the alchemical space. For converging the free energy surface in the alchemical space, both algorithms have mechanisms for decreasing the bias along the course of the simulation. In expanded ensemble, the Wang-Landau incrementor is modified by a scaling factor whenever the state visitation reaches a certain flatness criterion. In 1D well-tempered alchemical metadynamics, a bias factor is applied to enforce a continuous exponential decay of the Gaussian height, which leads to marginally smoother convergence compared to expanded ensemble or other similar free energy methods. There are a number of  alternative strategies that can also be pursued.~\cite{dama2014transition}.

In multi-dimensional alchemical metadynamics, introducing additional configurational CVs can further enhance the sampling of metastable states that might have been missed by methods that do not apply configurational biases. As the multi-dimensional biasing potentials can flatten out the free energy landscape in both configurational and alchemical space, the system would not get stuck in the phase space like scenarios B and C shown in Figure \ref{FES_scenarios}. This approach can be easily generalized to continuous alchemical space, but such a generalization is not explored in this study because methods such as $\lambda$-dynamics are not currently implemented in GROMACS. 

\subsection{Free energy calculations}
Theoretically, the free energy estimator for alchemical metadynamics is the same as the one used in any other metadynamics except that the CV vector is generalized with the introduction of the alchemical variable. Upon the deposition of the biasing potential $V(\bfv{\xi'})$ in alchemical metadynamics, the probability distribution sampled during the simulation is $\tilde{P}(\bfv{\xi'}) \propto \exp(-\beta(F(\bfv{\xi'}) + V(\bfv{\xi'})))$, where $\beta=1/k_{\text{B}}T$ is the inverse temperature. One of the possible options to recover the underlying free energy landscape $F(\bfv{\xi'})=-k_{\text{B}}T\ln P(\bfv{\xi'})$, is  to reweight the histogram by assigning an unbiasing weight $w(\bfv{\xi'})$ to each sample with the CV $\bfv{\xi'}$.~\cite{branduardi2012metadynamics} Such an unbiasing weight can be expressed as 
\begin{equation}
    w(\bfv{\xi'}) \propto \exp \left(\frac{V(\bfv{\xi'}, t_f)}{k_{\text{B}}T} \right)
\end{equation} where $t_f$ is the simulation length and $V(\bfv{\xi'}, t_f)$ is the total bias accumulated up to $t_f$. The maximum of $V(\bfv{\xi'}, t_f)$ over $\bfv{\xi'}$ is usually subtracted before taking the exponential to avoid overflow, which does not affect the normalized weights. More frequently, $V(\bfv{\xi'}, t_f)$ is replaced with $\bar{V}(\bfv{\xi'}, t_0)$, the total bias averaged over the time period from $t_0 = (1-f_a)t_f$ to $t_f$~\cite{micheletti2004reconstructing}, where $f_a$ is the fraction over which biases are averaged. Given that $t_f = t_0 + N\tau$, $\bar{V}(\bfv{\xi'}, t_0)$ can be written as 
\begin{equation}
    \bar{V}(\bfv{\xi'}, t_0) = \frac{1}{N+1}\sum_{i=0}^{N}V(\bfv{\xi'}, t_0 + i\tau)
\end{equation}
where $N$ is the number of Gaussians deposited from $t_0$ to $t_f$. Usually, $f_a$ is decided such that the bias potential applied during the period over which $\bar{V}$ is averaged is roughly stationary with time. To calculate $P(\bfv{\xi'})$ and its uncertainty (hence $F(\bfv{\xi'})$ and its uncertainty), we first discard the equilibrium phase during which the major free energy basins were being filled~\cite{bussi2020using}, with a truncation fraction of $f_{tr}$, which we here set to $1-f_{a}$. Then, we divide the remaining part of the simulation into blocks, for each of which we construct a weighted histogram of the CVs. Lastly, we calculate the free energy from the probability averaged over all the blocks, with the error of the free energy determined as the standard deviation of all bootstrap iterations in block bootstrapping. In practice, the uncertainty of the free energy is dependent on the number of blocks. We therefore calculated the uncertainty corresponding to different numbers of blocks ranging 20 up to 2000 and we report the maximum uncertainty. Alternatively, one could perform a separate simulation using a static bias potential and compute the weighting factors exclusively from the additional simulation, as it is done for instance in metadynamics with umbrella-sampling refinement~\cite{babin2006free}. Notice that this option is more expensive, as it require a separate simulation, but remove any potential systematic error due to the history-dependent nature of the metadynamics biasing potential.

\section{Methods}\label{methods}
We validated our implementation of alchemical metadynamics with free energy calculations for different test systems/alchemical processes with varying complexities and dimensions of the CV space. These range from decoupling an argon atom (Case 1) from water, or a model molecule composed of 4 interaction sites (Case 2, as shown in Figure \ref{sys2}) from water, to the methylation of a nucleobase and a duplex residue (Case 3, as shown in Figure \ref{sys3}). In the following subsections, we describe the simulation methods of different test systems, along with the details of the corresponding free energy calculations. All simulations were performed at 298K using GROMACS 2021.4. For metadynamics simulations, either a testing branch based on PLUMED 2.7 or PLUMED 2.8 was used, with no difference between the two versions except for the way of specifying relevant parameters. All files relevant to our simulations and test systems can be found in our project repository~\href{https://github.com/shirtsgroup/alchemical_MetaD}. 

\subsection{Case 1: Hydration of an argon atom}
As a sanity check for our implementation of alchemical metadynamics, we used 1D well-tempered alchemical metadynamics to calculate the solvation free energy of an argon atom, which was then compared with the result obtained from expanded ensemble given the similarity of the two methods. With System 1, the goal is to check if 1D alchemical metadynamics can accurately reproduce the free energy of a simple system calculated by expanded ensemble. 

\subsubsection{Preparation of simulation inputs}
The argon atom was solvated in a cubic box of length 2.4 nm and was energy-minimized by the steepest descent algorithm until the maximum force was lower than 100 kJ/mol/nm. The argon atom was modeled as a Lennard-Jones sphere with $\epsilon=0.996$ kJ/mol and $\sigma = 0.341$ nm. The system was then equilibrated in the NVT and then NPT ensembles, both for 200 ps. The reference temperature and pressure were maintained at 298 K and 1 bar by the velocity rescaling method~\cite{bussi2007canonical} and a Berendsen barostat~\cite{berendsen1984molecular}, respectively. Lastly, 5 ns of NPT MD simulation with Parrinello-Rahman barostat~\cite{parrinello1980crystal,parrinello1981polymorphic} keeping the pressure at 1 bar was performed, in which the cutoff distance for van der Waals interactions was specified as 0.9 nm. The PME (particle mesh Ewald) method~\cite{essmann1995smooth} was used with a switching function between 0.89 nm and 0.9 nm for efficient calculations of long-range electrostatic interactions. A spacing of 0.10 nm was used for the PME grids. The LINCS~\cite{hess1997lincs} algorithm was employed to constrain bonds involving hydrogens, with the highest order in the expansion of the constraint coupling matrix set as 12 and the number of iterative corrections set as 2. The configuration whose box volume was the closest to the average volume of the MD trajectory was extracted to serve as the input configuration for the expanded ensemble and alchemical metadynamics simulations, which were both performed in an NVT ensemble to avoid any potential issues with $\lambda$ dependence of pressure. Although this could lead to a slightly different estimate of the solvation free energy as compared to the one solved in the NPT ensemble, the objective is simply to compare two methods with the same alchemical process in the same ensemble. 

\subsubsection{Expanded ensemble simulation}
We divided our expanded ensemble calculations into two separate stages: an equilibration and a production stage. Both stages of simulations were performed in the NVT ensemble with 6 states for decoupling the van der Waals interactions between the argon atom and the water molecules. In the equilibration stage, we employed the Wang-Landau algorithm to adaptively estimate the weight for each alchemical state, with the initial Wang-Landau incrementor set as 0.5 $\text{k}_{\text{B}}\text{T}$, which is usually sufficient for relatively simple systems. The histogram that kept track of the state visitation was updated with the Metropolized-Gibbs Monte Carlo moves between all alchemical states, which were attempted every 10 integration steps. We adopted the default value of 0.8 for the cutoff for the flatness ratio $R$, which means that the histogram was considered flat only if all intermediate states had an $R$ value and its reciprocal larger than 0.8; For any state, $R$ is defined as the ratio between the count of the state and the average count of all states. Whenever the histogram was considered flat, the state counts were all reset to 0 and the Wang-Landau incrementor was scaled by a factor of 0.8. This process for updating weights was stopped when the incrementor fell below 0.001 $\text{k}_{\text{B}}\text{T}$. The equilibrated weights were then used as the frozen weights in the production simulation for 100 ns. For data analysis, we only considered the time series of the Hamiltonian obtained from the production stage. After truncating the non-equilibrium regime~\cite{chodera2016simple} and decorrelating the time series, we ran MBAR~\cite{shirts2008statistically} to compute the free energy difference between the coupled and uncoupled states, which is the solvation free energy of the system. The entire two-stage expanded ensemble procedure was done with 3 replicates. The final estimation of the solvation free energy is reported as the mean of the values obtained from the 3 replicates, with its uncertainty calculated as the error propagated from the bootstrapped uncertainty of the 3 replicates.

\subsubsection{1D alchemical metadynamics}
To compare with expanded ensemble, we adopted the same simulation length (100 ns), starting configuration, state transition scheme (Metropolized-Gibbs sampling), and coupling parameters of the same 6 states in 1D alchemical metadynamics, which is also done with 3 replicates. We set the initial height of the Gaussian biasing potential as 0.5 $\text{k}_{\text{B}}\text{T}$, which is the same as the initial Wang-Landau incrementor despite different weight updating schemes. While in alchemical metadynamics, the strides for Gaussian depositions and MC moves are decoupled and do not need to be the same, we set both strides as 10 integration steps to better compare with expanded ensemble, where the weights are always updated whenever a move is proposed. To accommodate such a fast pace for applying Gaussian biases, we set the bias factor as 50 to avoid an excessively fast decay in the Gaussian height, which could potentially slow down the compensation of the underlying free energy surface. The width of the Gaussian was set as 0.01 to avoid any overlap between the Gaussians deposited at different $\lambda$ values. Note that having such an overlap would not invalidate the method, but might have made the comparison to expanded ensemble less straightforward. For the solvation free energy calculation, we set the truncation fraction and the average fraction as 0.25 and 0.75, respectively. 1000 blocks were used in the histogram construction, which led to a block size of 75 ps and the largest uncertainty among the considered numbers of blocks. 200 bootstrap iterations were used in block bootstrapping. The mean and the propagated error of the free energy estimates from the 3 replicates were reported as the final results.

\subsection{Case 2: Hydration of a 4-site system}
\begin{figure}[ht]
    \centering
    \includegraphics[width=0.9\textwidth]{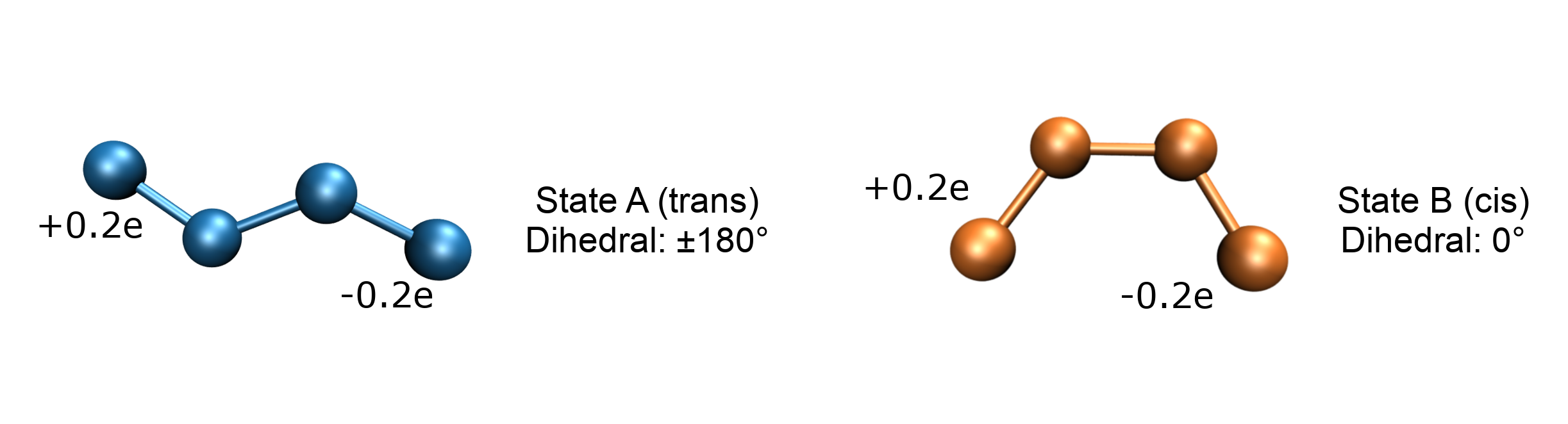}   
    \caption{The two torsional metastable states of System 2: A molecule composed of 4 interaction sites. The first and last atoms have charges of +0.2e and -0.2e, respectively, with the other two atoms uncharged.}
    \label{sys2}
\end{figure}

To demonstrate the advantages of introducing configurational CVs in alchemical metadynamics, we designed a fictitious molecule composed of 4 linearly bonded interaction sites as the second test system (see Figure \ref{sys2}). We placed opposite charges on the first and last ``atoms'' (+0.2e and -0.2e) and set the force constant for the only torsional angle as 60 kJ/mol so that the two torsional metastable states of the system were separated by a large free energy barrier (around 48.5 $\text{k}_{\text{B}}\text{T}$, see Figure S2A). This large free energy barrier poses a challenge of sampling both torsional metastable states to alchemical free energy methods that do not apply any configurational bias, such as expanded ensemble or 1D alchemical metadynamics. Such a challenge is useful for us to highlight the difference between methods with or without the application of configurational biases in free energy calculations. With this system, we performed 1D and 2D well-tempered alchemical metadynamics simulations starting from either metastable state to calculate the solvation free energy of the system. Since the cis and trans configurations have both different dipole moments and different effective volumes in solvent, configurational ensembles that are restricted to just one torsional well or the other will have different solvation free energies. The goal is to show that 2D alchemical metadynamics simulations starting from different torsional metastable states lead to statistically consistent estimates of the solvation free energy, which cannot be accomplished by 1D alchemical metadynamics due to restricted configurational sampling. 

\subsubsection{Preparation of simulation inputs}
After solvating the system in a dodecahedral box with 1 nm between the solute and box edges, we processed the system with the same procedure of energy minimization, NVT and NPT equilibration, and NPT MD simulation like the one used in the argon atom system. A structure that had a volume closest to the average NPT volume was taken as the input of a 5 ns NVT 1D metadynamics run that only biased the torsional angle of the system. This torsional metadynamics applied a Gaussian biasing potential every 500 integration steps, with a bias factor of 10. The width and initial height of the Gaussians were set as 0.5 rad and 1 $\text{k}_{\text{B}}\text{T}$, respectively. With this setup, the system was able to sample both torsional metastable states frequently in the torsional metadynamics (see Figure S2B), from which we generated the structures corresponding to the trans isomer (State A) and cis isomer (State B) for starting subsequent alchemical metadynamics elaborated in later sections. Notably, the parameters used in the torsional metadynamics do not need to be optimal as long as they are good enough for the simulation to generate reasonable starting configurations in the two isomer forms.

\subsubsection{Alchemical metadynamics}
For each torsional metastable state of the system, we started both 1D and 2D well-tempered alchemical metadynamics in an NVT ensemble, with 3 replicates for each, where the only torsional angle of the system was introduced in the 2D simulations as the second CV. The MC moves between alchemical states were proposed every 10 integration steps using the Metropolized-Gibbs MC scheme. All simulations were performed for 200 ns and adopted a Gaussian deposition stride of 500 steps. 20 alchemical intermediate states were used to decouple the van der Waals interactions and Coulombic interactions. 1D alchemical metadynamics started with a Gaussian height of 1 $\text{k}_{\text{B}}\text{T}$. A general metadynamics rule of thumb is that the bias factor should be approximately $\Delta G/k_{\text{B}}T$, where $\Delta G$ is the height of the free energy barrier to cross~\cite{WTMetaD}.  The bias factor was thus set as 60 for the 1D alchemical metadynamics. This choice was guided by the fact that typical free-energy differences in the alchemical space are around 50 kT (see Figure S2A) and, given the expected equilibrium distribution of well-tempered metadynamics, this bias factor allows a reasonably flat histogram of lambda. On the other hand, 2D alchemical metadynamics used a Gaussian height of 2 $\text{k}_{\text{B}}\text{T}$, and a bias factor of 120 for flattening the deeper free energy basins in the 2D phase space. The widths of the Gaussian along the alchemical direction (for 1D and 2D simulations) were set as 0.01, while the Gaussian width in the torsional dimension (for the 2D simulations) was set as 0.5 rad. While the adopted Gaussian width along the torsional direction is slightly larger than the typically suggested value of 0.35 rads for biasing torsions, it has also been shown~\cite{branduardi2012metadynamics} that wider Gaussians could fill free energy basins faster. For all simulations, we specified a truncation fraction of 0.3 and an average fraction of 0.7. For each simulation, the block size that led to the largest uncertainty was adopted for histogramming in free energy calculations and 200 bootstrap iterations were used in block bootstrapping. For each of the 4 kinds of calculations (1D or 2D simulations starting from either State A or State B), the final results were calculated as the mean and the propagated error of the free energy estimates from the 3 replicates.

\subsection{Case 3: Adenosine in its isolated form and in a duplex}
\begin{figure}[ht]
    \centering
    \includegraphics[width=0.9\textwidth]{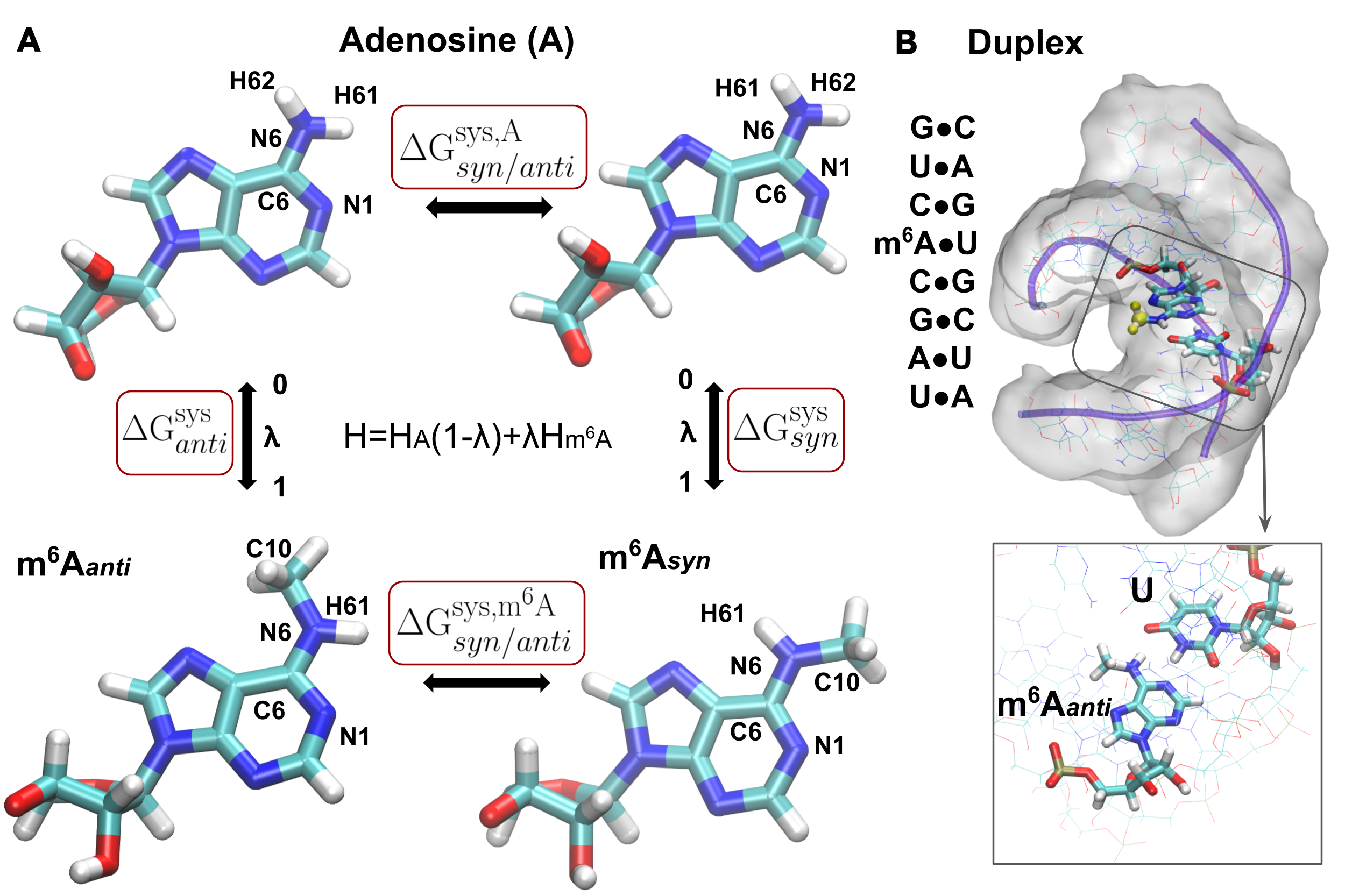}   
    \caption{(A) The 4 considered states of the alchemical transformation of A into m$^6$A. Isomers are characterised by the value of the torsional angle defined by atoms N1-C6-N6-H62 or N1-C6-N6-C10. The isomers are indistinguishable in the adenine case, so $\Delta G^{\text{sys,A}}_{syn/anti}=0$. On the other hand, in m$^6$A the position of the methyl group defines the states \emph{anti} and \emph{syn}. The former is the most favoured for the paired m$^6$A in a duplex, while the latter is the most favoured for the isolated nucleoside. (B) The 8 base-pairs duplex considered in this work, shown in the case of methylated adenosine in \emph{anti} state.}
    \label{sys3}
\end{figure}

As a final example, we considered the free energy calculations associated to the modification of adenosine (A) to N6-methylated adenosine (m$^6$A). m$^6$A is the most widespread post-transcriptional modification of RNA~\cite{gilbert2016messenger}. The methyl group can be either in \emph{syn} or \emph{anti} configuration, distinguished by a torsional angle $\eta$ (see Figure \ref{sys3}). When m$^6$A is isolated (nucleoside) or in a single-stranded region, the \emph{syn} state is more stable than the \emph{anti} state. The reverse is true when m$^6$A is in a duplex. This results in an effective duplex destabilization~\cite{roost2015structure}. The barriers associated with the $\eta$ angle are relevant to the kinetics of hybridization~\cite{liu2021quantitative}. In a previous paper~\cite{piomponi2022molecular}, we studied this system using standard Hamiltonian replica exchange simulations and reparametrized partial charges so as to fit thermodynamic data~\cite{roost2015structure, kierzek2022secondary}. Specifically, we separately simulated the \emph{syn}- and \emph{anti}- conformations. We here show how to use alchemical metadynamics to recover the same information in a single simulation, and additionally obtaining information about the isomerization barrier.

\subsubsection{Preparation of simulation inputs}
All the setups have been described extensively in Piomponi et al.,~\cite{piomponi2022molecular} and are available on  \href{https://zenodo.org/record/6498021}{Zenodo}. We here consider the conversion of A to m$^6$A in an isolated nucleoside (system A1 in Ref. \cite{piomponi2022molecular}) and in a RNA duplex (system A2 in Ref. \cite{piomponi2022molecular}, only the duplex system). The GROMACS input files are identical to those used in our previous work, except that here the $\lambda$ ladder is sampled with the Metropolized-Gibbs algorithm with attempted moves spaced with 100 integration steps. For the simulations reported in this work, we used the parametrization of m$^6$A charges reported as fit\_A in Ref. \cite{piomponi2022molecular}.

\subsubsection{Alchemical metadynamics}
Similarly to the other two systems, we use metadynamics to flatten the sampling along both the alchemical $\lambda$ state and along a physical collective variable. For this system, we tested a modified setup where we apply two concurrent metadynamics~\cite{gil2015enhanced}. The first metadynamics process is one-dimensional and acts only along the alchemical variable. Since the free energy differences along this non-physical variable can be very large, we use a large bias factor ($\gamma=100$). The second, simultaneous, metadynamics process is two-dimensional and acts both on the alchemical variable and on $\eta_{\text{avg}}$, an averaged torsional angle elaborated in the next section. Since the barriers along $\eta_{\text{avg}}$ are smaller, this second metadynamics is performed with a lower bias factor ($\gamma=10$). The overall bias potential acting on the system can thus be written as
\begin{equation}
    V_{\text{tot}}(\eta_{\text{avg}},\lambda) = V_1(\lambda) + V_2(\eta_{\text{avg}},\lambda)
\end{equation}
where $V_1$ and $V_2$ are the Gaussian biases added during the one-dimensional metadynamics and the two-dimensional metadynamics, respectively. This combined bias potential can be directly used for reweighting as discussed above. Notably, at variance with the work by Gil-Ley et al.,~\cite{gil2015enhanced} where a large number of collective variables were concurrently biased, thus requiring a replica ladder to obtain unbiased populations, by using only two variables as we do here a direct reweighting is sufficient. Notably, a similar issue appears when simultaneously biasing the total energy of a solvated system and solute-dependent CVs. In this case, indeed, two separate metadynamics, possibly with different bias factors, can be applied fruitfully. This was done for instance in Ref. ~\cite{deighan2012efficient}, though in a sequential and not self-consistent procedure.  The protocol is also related to the one proposed by Chipot and Leli\`evre~\cite{chipot2011enhanced}, although it is here applied (a) in metadynamics context and (b) combining potentials in 1D and 2D with the alchemical CV shared among the two biases.

Metadynamics simulations were run for 60 ns, with Gaussians of initial height 12 kJ/mol, for $V_1$, and 1.2 kJ/mol, for $V_2$, deposited every 500 steps. The Gaussian width along the $\eta_{\text{avg}}$ variable was chosen to be 0.35 rad. The 2D free energy surface was computed directly from the bias potentials, while the 1D profile was reconstructed using reweighting. Free energy differences and their statistical errors were computed by reweighting a second 160 ns-long simulation where the bias potentials were kept constant. In the case of this calculation, as has also been observed anecdotally in other cases, using a static bias resulted in slightly more statistically robust free energy differences. For a comparison between the cases using dynamic or static bias, please refer to the supporting information, or more specifically, Figure S1.

\subsubsection{Choice of the configurational collective variable}
One critical issue in this system is the proper choice of the configurational collective variable. In the first attempt, we used the torsional angle defined as the torsion identified by atoms N1-C6-N6-C10 (see Figure~\ref{sys3}). This choice was found to be suboptimal. In the production runs, we used as a biased variable a mean torsion obtained by averaging the three torsions identified by atoms N1-C6-N6-C10,  N1-C6-N6-H61, and N1-C6-N6-H62. The average was computed as the arctangent of the sine and cosine averages. These three torsions are coupled by an improper torsion that maintains the group C10, N6, H61, and H62 planar, but this torsion is not sufficiently stiff to maintain the consistency between the three torsions when enforcing the barrier crossing. When biasing the average, a diffusive behavior of the biased CV was obtained (Figure S6B).
Specifically, with the torsions N1-C6-N6-C10 ($\eta_{\text{C10}}$), N1-C6-N6-H61 ($\eta_{\text{H61}}$), and N1-C6-N6-H62 ($\eta_{\text{H62}}$), the average is computed as 
\begin{equation}
    \eta _{\text{avg}}=\textrm{atan2}\left(\frac{\sin(\eta_{\text{C10}})+\sin(\eta_{\text{H61}}+\pi)+\sin(\eta_{\text{H62}})}{3},\frac{\cos(\eta_{\text{C10}})+\cos(\eta_{\text{H61}}+\pi)+\cos(\eta_{\text{H62}})}{3}\right)
\end{equation}
Where atan2 is is the two argument arctan function, defined as the angle between the positive $x$-axis and the vector $(x,y)$; it is equal to $\arctan(y/x)$ when $x> 0$, but involves corrections of $\pm\pi$ when $x \le 0$.  We also note that $\eta_{\text{H61}}$ must be shifted by 180 degrees when taking the average.

\subsubsection{Free energy calculations}
For this system, we are interested in calculating of the following three relative free energy differences: $\Delta \Delta G^{\text{ns}}_{syn/anti}$, $\Delta \Delta G^{\text{dup}}_{syn/anti}$, and $\Delta \Delta G^\text{dup/ns}_{syn+anti}$, where the first two denote the difference in the methylation free energy between the transformation processes that lead to a \emph{syn} or \emph{anti} m$^6$A, in the isolated nucleoside (ns) and in the duplex (dup), respectively. They can be calculated by taking the difference between the free energy differences of interest, namely, 

\begin{equation}
    \Delta \Delta G^{\text{ns}}_{syn/anti}= \Delta G^{\text{ns}}_{anti}- \Delta G^{\text{ns}}_{syn} 
    \label{DG_ns}
\end{equation} 
\begin{equation}
    \Delta \Delta G^{\text{dup}}_{syn/anti}= \Delta G^{\text{dup}}_{anti}- \Delta G^{\text{dup}}_{syn} 
    \label{DG_dup}
\end{equation} 
The same set of free energy differences can be used to calculate $\Delta \Delta G^\text{dup/ns}_{syn+anti}$, the relative methylation free energy between the nucleoside and the duplex systems considering both \emph{syn} and \emph{anti} conformations: 
\begin{equation}
\Delta \Delta G^\text{dup/ns}_{syn + anti}=\Delta G^{\text{dup}}_{syn + anti} - \Delta G^{\text{ns}}_{syn + anti}
\label{DDG_3}
\end{equation}
with
\begin{equation}
\Delta G^{\text{ns}}_{syn + anti} = -\frac{1}{\beta}\ln(\exp(-\Delta G^{\text{ns}}_{syn}) + \exp(-\Delta G^{\text{ns}}_{anti}))
\label{methylation_1}
\end{equation}
\begin{equation}
\Delta G^{\text{dup}}_{syn+anti} = -\frac{1}{\beta}\ln(\exp(-\Delta G^{\text{dup}}_{syn}) + \exp(-\Delta G^{\text{dup}}_{anti}))
\label{methylation_2}
\end{equation}
In Equations \ref{DG_ns}, \ref{DG_dup}, \ref{methylation_1} and \ref{methylation_2}, $\Delta G^{\text{ns}}_{syn}$, $\Delta G^{\text{ns}}_{anti}$, $\Delta G^{\text{duplex}}_{syn}$, and $\Delta G^{\text{duplex}}_{anti}$ are the free energy differences of converting an adenosine into a \emph{syn} m$^6$A or \emph{anti} m$^6$A in either the isolated form or the duplex, each of which can be calculated from a separate alchemical simulation at fixed rotameric state. For example, in the work by Piomponi et al.,~\cite{piomponi2022molecular} four Hamiltonian replica exchange simulations were performed to estimate the three relative free energy differences of interest ($\Delta \Delta G^{\text{ns}}_{syn/anti}$, $\Delta \Delta G^{\text{dup}}_{syn/anti}$, and $\Delta \Delta G^\text{dup/ns}_{syn+anti}$).

However, using alchemical metadynamics, we can sample both rotamers in a single simulation methylating the adenosine. Thus, $\Delta G^{\text{sys}}_{syn}$, $\Delta G^{\text{sys}}_{anti}$, with $^\text{sys}$ being either $^\text{ns}$ or $^\text{dup}$, can be directly obtained from a single alchemical metadynamics simulation. Given the access to all metastable states in the alchemical and configurational space, we can calculate free energy differences with more flexibility by considering ratios of partition functions corresponding to different states. For example, with alchemical metadynamics, we can calculate $\Delta \Delta G^{\text{ns}}_{syn/anti}$ and $\Delta \Delta G^{\text{dup}}_{syn/anti}$ as follows, instead of using Equations \ref{DG_ns} and \ref{DG_dup}:
\begin{equation}
    \Delta \Delta G^{\text{ns}}_{syn/anti} = \Delta G^{\text{ns, m}^6{\text{A}}}_{syn/anti} = -\frac{1}{\beta} \ln \left( \frac{\sum_{i \in anti} e^{\beta V^{\text{ns}}_{\text{tot}}(\eta_i, \lambda=1) }}{\sum_{i \in syn} e^{\beta V^{\text{ns}}_{\text{tot}}(\eta_i, \lambda=1)}}\right)
\end{equation}

\begin{equation}
    \Delta \Delta G^{\text{dup}}_{syn/anti} = \Delta G^{\text{dup, m}^6{\text{A}}}_{syn/anti} = -\frac{1}{\beta} \ln \left( \frac{\sum_{i \in anti} e^{\beta V^{\text{dup}}_{\text{tot}}(\eta_i, \lambda=1) }}{\sum_{i \in syn} e^{\beta V^{\text{dup}}_{\text{tot}}(\eta_i, \lambda=1)}}\right )
\end{equation}

$\Delta G^{\text{ns, m}^6{\text{A}}}_{syn/anti}$ and $\Delta G^{\text{dup, m}^6{\text{A}}}_{syn/anti}$, which are the free energy differences between the two rotamers in the nucleoside and in the duplex, respectively, are not available in Hamiltonian replica exchange but in alchemical metadynamics. Similarly, $\Delta G^{\text{ns}}_{syn + anti}$ and $\Delta G^{\text{ns}}_{syn + anti}$ can be calculated as follows:

\begin{equation}
    \Delta G^{\text{ns}}_{syn + anti} = -\frac{1}{\beta} \ln \left( \frac{\sum_{i \in syn+anti} e^{\beta V^{\text{ns}}_{\text{tot}}(\eta_i, \lambda=1) }}{\sum_{i \in syn+anti} e^{\beta V^{\text{ns}}_{\text{tot}}(\eta_i, \lambda=0)}}\right )
\end{equation}

\begin{equation}
    \Delta G^{\text{dup}}_{syn + anti} = -\frac{1}{\beta} \ln \left( \frac{\sum_{i \in syn+anti} e^{\beta V^{\text{dup}}_{\text{tot}}(\eta_i, \lambda=1) }}{\sum_{i \in syn+anti} e^{\beta V^{\text{dup}}_{\text{tot}}(\eta_i, \lambda=0)}}\right )
\end{equation}
so that $\Delta \Delta G^\text{dup/ns}_{syn + anti}$ can be calculated using Equation \ref{DDG_3}. In Case 3, the goal is to compare the three free energy differences obtained from alchemical metadynamics with the values recovered from Hamiltonian replica exchange reported in the work by Piomponi et al~\cite{piomponi2022molecular}.

\section{Results and discussion}
\subsection{Case 1: An argon atom}
With the weights fixed at the values equilibrated by the Wang-Landau algorithm, the solvation free energy of the argon atom estimated by expanded ensemble with MBAR was $-3.275$ $\text{k}_{\text{B}}\text{T}$, with an uncertainty as small as 0.016 $\text{k}_{\text{B}}\text{T}$ owing to sufficiently even state visitation (see Figure S3A). This estimation is statistically consistent with the one obtained from the 1D alchemical metadynamics, which was $-3.284$ $\pm$ 0.010 $\text{k}_{\text{B}}\text{T}$. Notably, the essential difference in the weight updating approaches between the two methods makes it infeasible to directly compare the performance of the methods. In expanded ensemble, the incrementor decreases in a step-wise manner and the same value is applied to all intermediate states. On the other hand, the Gaussian height in well-tempered metadynamics continuously decays with the amount of biases that have been deposited in the state being sampled, which means that the potential energies of different states are elevated with different amounts depending on how frequently the states have been visited. If we had adopted a stricter, rather than a typical criterion for histogram flatness, our state visitation of expanded ensemble would have been more even. This would also result in a lower uncertainty. With the chosen parameters in our case, though, the expanded ensemble still reached a low uncertainty. More importantly, the data collected from the two simulations is sufficient to indicate that 1D alchemical metadynamics free energy calculations yield results that are equivalent within statistical significance to the expanded ensemble results.

\subsection{Case 2: A molecule composed of 4 interaction sites}
\begin{figure}[ht]
    \centering
    \includegraphics[width=\textwidth]{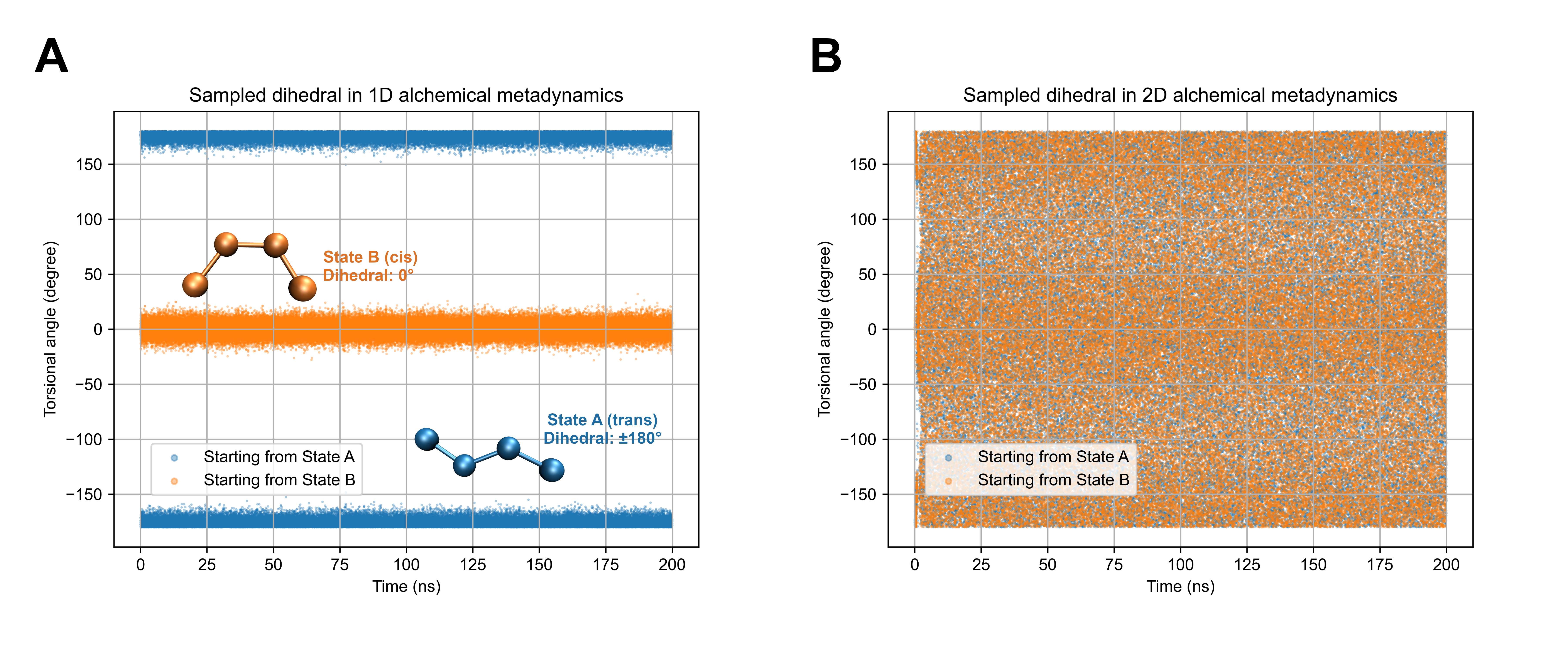}   
    \caption{The sampled torsional angle in (A) 1D alchemical metadynamics and (B) 2D alchemical metadynamics as a function of time. As can be seen in the figure, the sampling in the torsional space is restricted in 1D alchemical metadynamics, but essentially complete in its 2D analog.}
    \label{dihedral_A_B}
\end{figure}

\begin{figure}[H]
    \centering
    \includegraphics[width=0.7\textwidth]{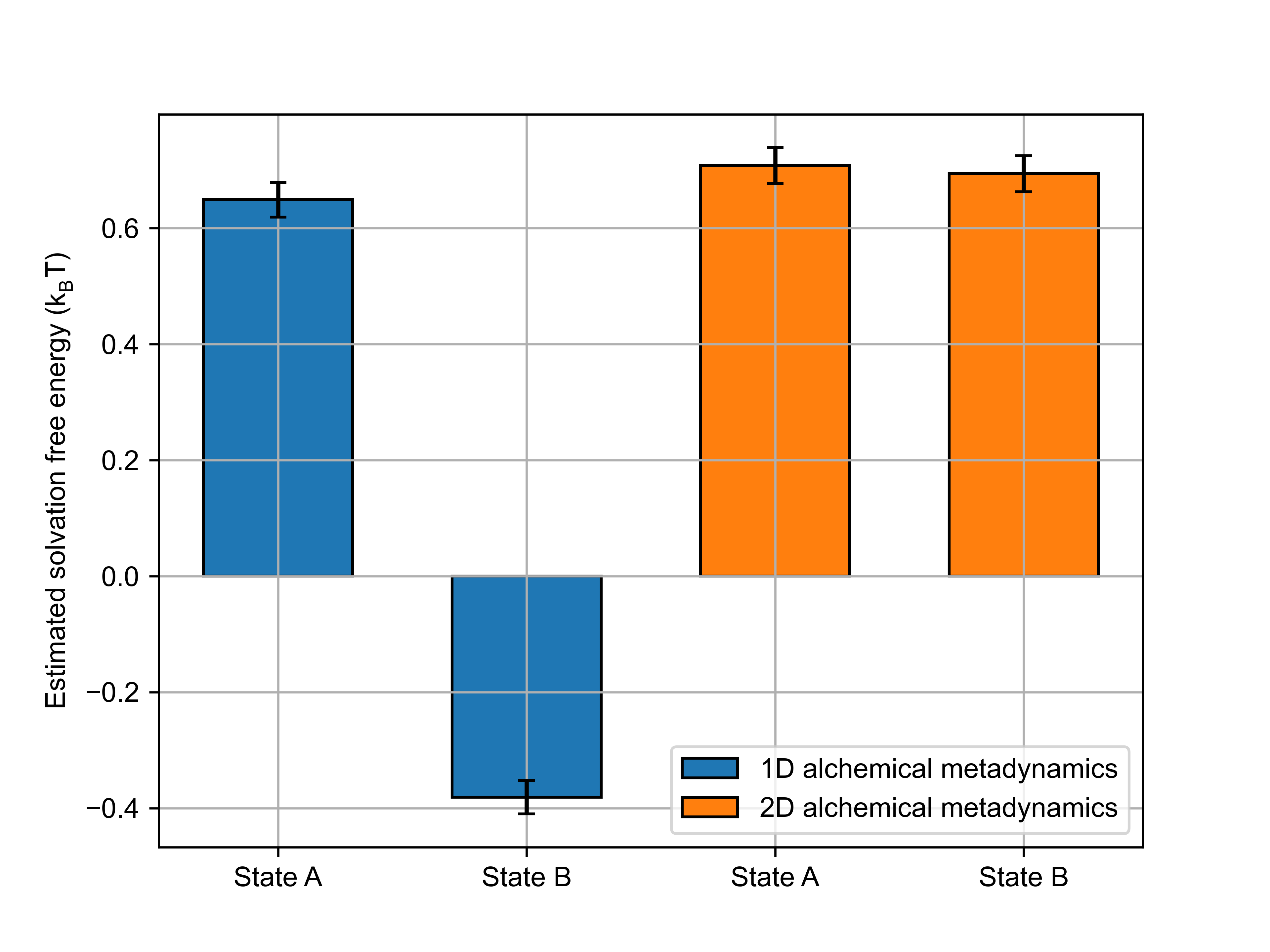}
    \caption{The solvation free energies estimated by 1D and 2D alchemical metadynamics starting from either states. 1D alchemical metadynamics simulations starting from different torsional metastable states led to statistically different estimations of the solvation free energy, while the values estimated by the 2D simulations are statistically consistent with each other.}
    \label{sys2_df_results}
\end{figure}

The purpose of the second system is to demonstrate the difference in the configurational sampling between alchemical metadynamics with or without the introduction of configurational bias. As a consequence, in 1D alchemical metadynamics starting from either torsional metastable states, the accumulation of one-dimensional alchemical biases allowed the system to freely sample all the intermediate states (see Figure S4). However, such biases did not facilitate the compensation of the free energy wells along the torsional direction. With the lack of direct biases in the torsional direction, the interconversion of the two isomers became the slowest degree of freedom that trapped the system. Accordingly, it can be seen in Figure \ref{dihedral_A_B}A that 1D alchemical metadynamics failed to sample both metastable states regardless of which torsional state the simulation was initialized in. This insufficient sampling of the torsional space caused the dependence of the estimated solvation free energy on the starting torsional metastable state. Specifically, the solvation free energies estimated by 1D alchemical metadynamics starting from State A and State B were 0.649 $\pm$ 0.030 $\text{k}_{\text{B}}\text{T}$ and -0.381 $\pm$ 0.029 $\text{k}_{\text{B}}\text{T}$ (see Figure \ref{sys2_df_results}), respectively. As either simulation failed to account for the potential energy contribution of the other metastable state, the free energy estimates were statistically inconsistent with each other. 

\begin{figure}[H]
    \centering
    \includegraphics[width=\textwidth]{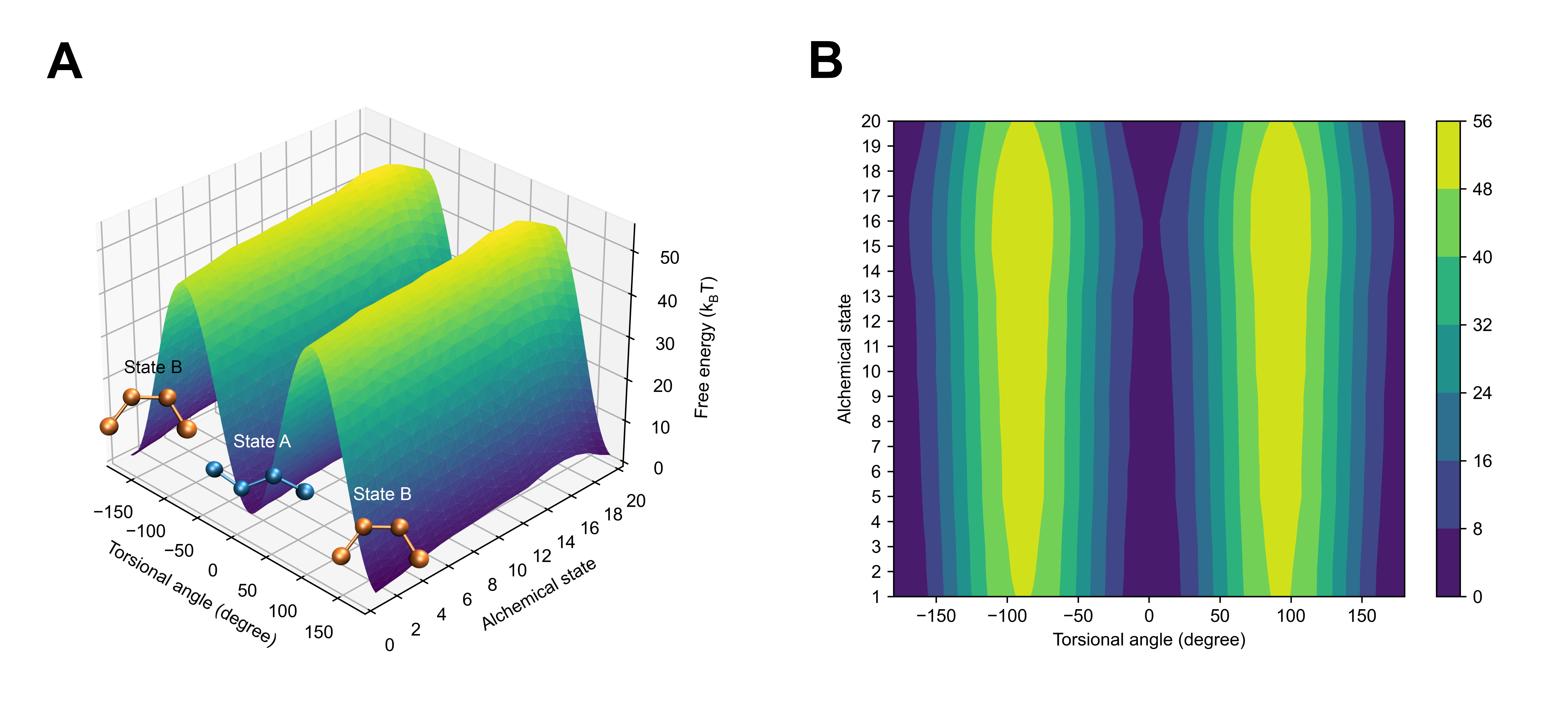}   
    \caption{(A) The average of the 2D free energy surfaces obtained from the two 2D alchemical metadynamics simulations starting from State A and State B. (B) The average of the contour plots obtained from the 2D alchemical metadynamics simulations starting from State A and State B.}
    \label{sys2_fes_contour}
\end{figure}

By contrast, the 2D Gaussian biasing potentials in both 2D alchemical metadynamics simulations flattened out the free energy surface both along the alchemical and torsional directions simultaneously, so in both cases, the system was able to sample the alchemical and torsional space exhaustively (see Figure S5 and Figure \ref{dihedral_A_B}B). This comprehensive and even sampling along all the slow degrees of freedom led to statistically consistent solvation free energy estimations from State A (0.708 $\pm$ 0.031 $\text{k}_{\text{B}}\text{T}$) and State B (0.694 $\pm$ 0.031 $\text{k}_{\text{B}}\text{T}$), as shown in Figure \ref{sys2_df_results}. In addition, the 2D free energy surface of the system can be accurately recovered from either of the two cases. Figure \ref{sys2_fes_contour}A shows the an averaged 2D free energy surfaces obtained from one of the 3 replicates, which was calculated by averaging the 2D free energy surfaces obtained from the 2D alchemical metadynamics starting from the two torsional metastable states. Figure \ref{sys2_fes_contour}B shows the contour plot corresponding to the averaged 2D free energy surface in \ref{sys2_fes_contour}A. From the 2D free energy surface, it is clear that the alchemical variable is orthogonal to the torsional angle of the system, which explains the ineffectiveness of the one-dimensional alchemical bias in torsional sampling. Notably, this is exactly Scenario B in Figure \ref{FES_scenarios} that can fail alchemical free energy methods that do not apply configurational biases. Therefore, the success in System 2 verified the usage of 2D alchemical metadynamics in overcoming the extensive free energy barrier present in a certain configurational CV direction. 

\subsection{Case 3: Adenosine methylation in its isolated form and in a duplex}
\begin{figure}[H]
    \centering
    \includegraphics[width=\textwidth]{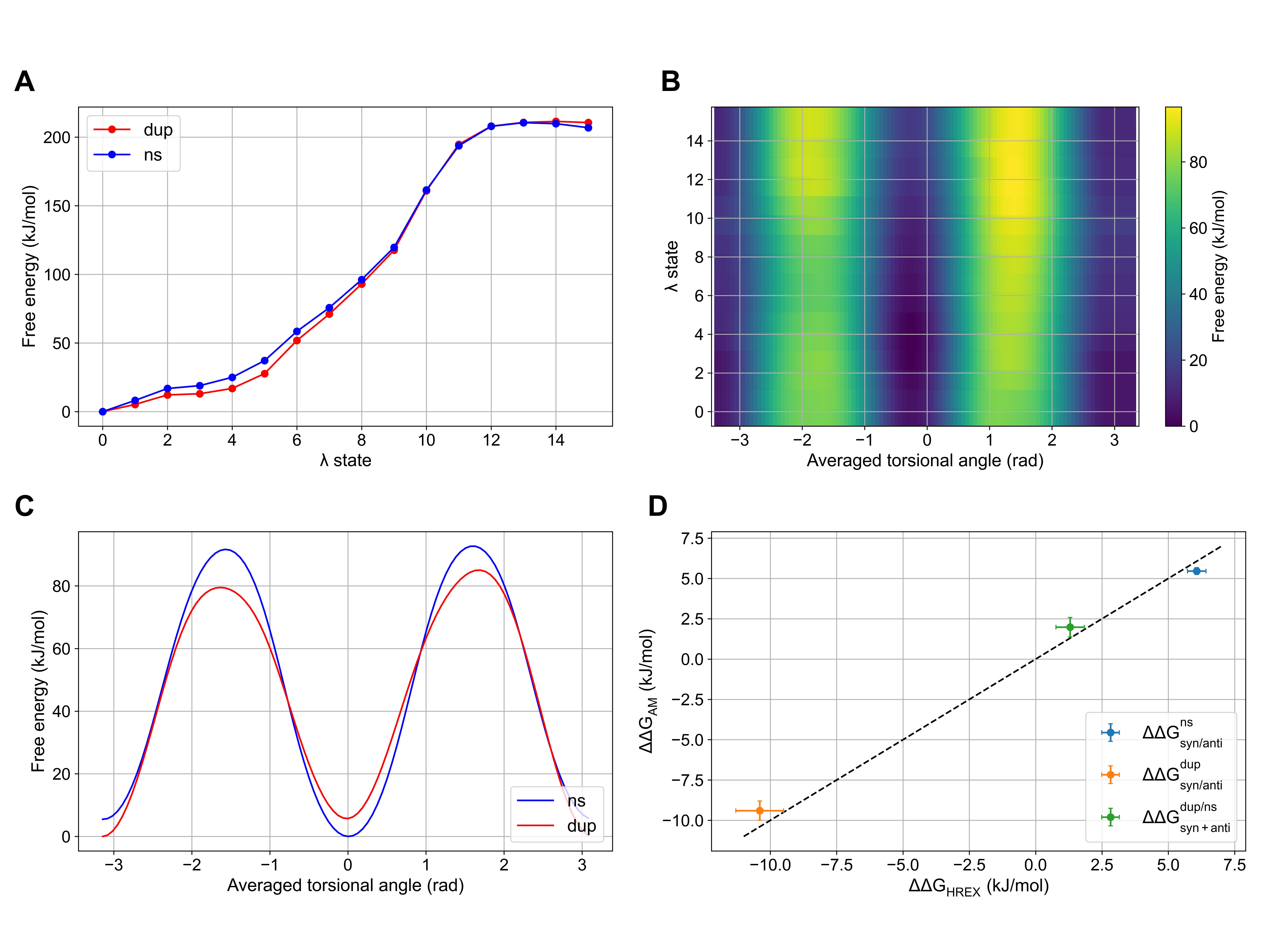}   
    \caption{(A) The free energy profile along the state index for the RNA duplex and for the isolated nucleoside. (B) Residual free energy surface along the state index and the averaged torsional angle for the RNA duplex (C) The free energy computed as a function of $\eta_{\text{avg}}$ at fixed $\lambda=1$, both for the RNA duplex (red) and for the m$^6$A nucleoside (blue) (D) Comparison of $\Delta \Delta G$ obtained with alchemical metadynamics (AM) and with Hamiltonian replica exchange (HREX) from Piomponi et al., 2022~\cite{piomponi2022molecular}, with their respective statistical errors.}
    \label{m6A_results}
\end{figure}

The free energy profile along the $\lambda$ state index is computed using reweighting and reported in Figure \ref{m6A_results}A. The large difference observed is non-physical and depends on the relative definitions of the A and m$^6$A force field parameters. The 2D surface as a function of the $\lambda$ state index and the averaged torsional angle $\eta_{\text{avg}}$ is computed using the usual relationship between bias and free energy~\cite{WTMetaD}, and then subtracting the Boltzmann-averaged free energy along the $\lambda$ state index, and is reported in Figure \ref{m6A_results}B. We notice that the residual dependence of the free energy on $\lambda$ depends on the fact that barriers on $\eta_{\text{avg}}$ change when $\lambda$ is changed. The profiles along $\eta_{\text{avg}}$ were computed using the relationship between the bias and the free energy and are also shown Figure \ref{m6A_results}C. Notably, this approach allows free energy profiles along the biased variable to be obtained simultaneously with alchemical differences. These profiles show that the \emph{syn} conformation (central basin) is favored in the m$^6$A nucleoside, whereas the \emph{anti} conformation (lateral basins) is favored in the duplex. The final $\Delta \Delta G$'s, which represent the amount by which the methylation disfavors the duplex, are consistent with those reported in Ref.~\cite{piomponi2022molecular} within the respective statistical errors (Figure \ref{m6A_results}D).

Importantly, as it is common in all methods based on biasing collective variables, the choice of the collective variable is critical. In the simulations reported above, a diffusive behavior was observed in the collective variable after the main basins were filled. We report in Figure S6A preliminary results obtained with a suboptimal variable where a diffusive behavior along $\eta$ was not obtained. Importantly, even though the exploration of $\lambda$ is guaranteed by the one-dimensional metadynamics, the inclusion of $\lambda$ in the two-dimensional metadynamics allows to effectively reconstruct free energies along $\eta_{\text{avg}}$ that are depending on $\lambda$.

In the third case reported in this paper, the alchemical simulation of conversion from A to m$^6$A is an interesting physical example because it shows that alchemical metadynamics gives simultaneous access to free energy barriers for both the two end systems. Whereas this result could have been obtained performing two separate metadynamics simulations, being able to use a single simulation has substantial advantages. First, it makes sure that other possibly slow degrees of freedom are sampled consistently in the two end states, making differential results more reliable. For instance, if the isomerization barrier was affected by binding with another molecule present in the simulation box, the dynamics of $\lambda$ would have ensured binding to be equally represented in the A and m$^6$A states. Second, in cases where the conformational transitions are better described by the physical CV in one of the states with respect to the other state, thus resulting in more transitions in one of the end states when compared to the other, having a single simulation would enable the ensemble of the slower state to benefit from the enhanced ergodicity in the faster state. These benefits could also be obtained by combining metadynamics with Hamiltonian replica exchange along the alchemical variable, however at the price of higher computational cost and less flexibility in the setup.

The combination of one-dimensional and two-dimensional bias potentials allows simultaneously (a) flattening of the large artificial free-energy difference along the alchemical variable and (b) compensating of the torsional barriers effectively, considering the fact that the precise profiles depend on the alchemical variable. The two potentials can be constructed using different bias factor coefficients so as to optimize their capability to explore the two profiles. This idea might be also exploited in different contexts, whenever one wants to simultaneously facilitate transitions over a large free energy barrier (e.g., a chemical reaction) and, at the same time, smooth residual barriers on softer degrees of freedom.

\section{Conclusion}
In this study, we proposed alchemical metadynamics, which expanded the configurationally defined sampling space allowed in traditional metadynamics with an additional alchemical sampling direction. Alchemical metadynamics is most useful when the CV space is multi-dimensional, including both alchemical and configurational sampling. With the configurational bias, it encourages the system to escape from configurational metastable subspace that could have easily trapped the system. It retains the advantages of traditional alchemical free energy methods, but also enables higher flexibility in sampling rough free energy surfaces. 

With different test systems and alchemical processes, we showed that 1D alchemical metadynamics had at least comparable performance as expanded ensemble simulations, and was able to accurately calculate the solvation free energy of an argon atom. We also showed that 2D alchemical metadynamics could eliminate the dependence of free energy calculations on the starting metastable state due to restricted configurational sampling with system 2. With system 3, we demonstrated that 2D alchemical metadynamics eliminated the need to perform multiple Hamiltonian replica exchange simulations to estimate the relative methylation free energy of the adenosine systems and simultaneously reconstructed the free energy profile along the biased torsional angle. More importantly, the success in both Cases 2 and 3 manifests the usage of alchemical metadynamics in overcoming challenges that can frustrate traditional alchemical free energy methods that do not bias configurational CVs. We conclude that alchemical metadynamics is promising in enhancing sampling in challenging systems, such as highly flexible protein-peptide binding complexes, or protein-nucleic acid binding complexes. The method can be trivially combined with more sophisticated algorithms, such as path collective variable~\cite{branduardi2007b} tICA~\cite{m2017tica}, SGOOP~\cite{tiwary2016spectral}, RAVE~\cite{ribeiro2018reweighted}, or other similar machine learning methods~\cite{rohrdanz2011determination, sultan2018automated, mccarty2017variational, chen2018molecular, wang2019past, wehmeyer2018time}.

\section*{Author Contributions}
W.-T.H., M.R.S. primarily conceptualized the project and designed the methodology, with additional contributions from P.T.M and G.B.; P.T.M. implemented the sampling methods in PLUMED and GROMACS, with contributions from G.B. Experiments were performed by W.-T.H. and V.P.; experiments were analyzed and validated by W.-T.H. and V.P. with contributions from M.R.S.; W.-T.H wrote the original manuscript draft; editing and review of the manuscript was done by M.R.S., G.B. and P.T.M. M.R.S. supervised the project and obtained resources.

\begin{acknowledgement}
This study was supported by the grant from National Science Foundation (OAC-1835720) (W.-T.S and M.R.S) and an MolSSI Software Fellowship (P.T.M.). MolSSI is funded by NSF CHE-2136142.  The computational work done in this publication used resources provided from the Extreme Science and Engineering Discovery Environment (XSEDE), which is supported by National Science Foundation grant number ACI-1548562. Specifically, it used the Bridges-2 system, which is supported by NSF  ACI-1928147, located at the Pittsburgh Supercomputing Center (PSC). 

\end{acknowledgement}



\input{main.bbl}

\clearpage
For Table of Contents Only
\begin{figure}[H]
    \centering
    \includegraphics[width=\textwidth]{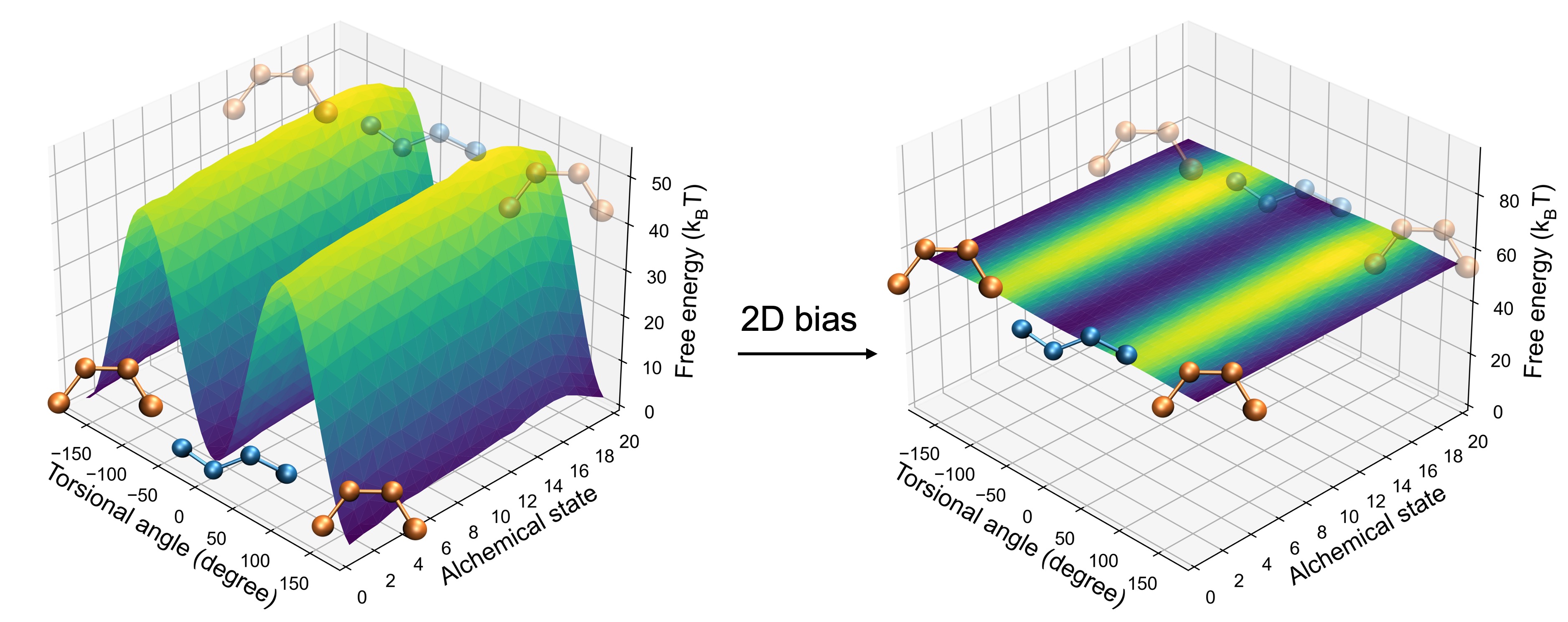}   
    \caption{}
\end{figure}
\end{document}


\section{Comparison of methylation free energy calculations with dynamic and static biases}
As a supplementary information, we show the free energy calculations with dynamic bias for the nucleotide and duplex systems. These calculations are done in comparison with the free energy differences computed with static bias presented in the main text. Specifically, simulations at dynamic bias were elongated up to 160 ns. For analysis, the first 60 ns were discarded, and the bias averaged over the remaining 100 ns was used to compute weights. Different numbers of blocks ranging 2 to 1000 were used to construct histograms in block boostrapping (200 iterations) and the largest uncertainty is reported. 

The figure below shows that with dynamic bias, the free energy estimates are more precise (lower statistical errors). This is mostly likely  attributable to the fact that the sampling in the CV space is more diffusive in these systems with dynamically updated weights. However, free energy estimates computed with dynamic bias are less accurate, i.e. the results differ more in alchemical metadynamics than in the case of Hamiltonian replica exchange (HREX). This is probably caused by the dynamic bias adding some small amount of history-dependent blurring. 

To further demonstrate the lower accuracy of the dynamic bias computation, the free energy difference ($\Delta G^{\text{dup, A}}_{syn/anti}$) between the two conformations of adenosine shown in Figure 3A in the main text is calculated. In the work by Piomponi et al.,~\cite{piomponi2022molecular} this value was assumed to be 0 because of the symmetry of the hydrogen atoms H61 and H62. Also, HREX used in the previous work does not have the access to the free energy landscape along the biased torsion, so the relative error is not given for the HREX case. In alchemical metadynamics, $\Delta G^{\text{dup, A}}_{anti/syn}$ was calculated as follows: 
\begin{equation}
    \Delta G^{\text{dup, A}}_{syn/anti}=-\frac{1}{\beta} \ln \left( \frac{\sum_{i \in anti} e^{\beta V^{\text{dup}}_{\text{tot}}(\eta_i, \lambda=0) }}{\sum_{i \in syn} e^{\beta V^{\text{dup}}_{\text{tot}}(\eta_i, \lambda=0)}}\right )
\end{equation} 

For most systems, the general understanding is that using plain metadynamics instead of doing the two-step procedure is better~\cite{bussi2020using}. It is likely the result is system dependent and related to the fact that even without a dynamic bias we can see many of transitions, thus a reasonable statistical error. In this way, we are clearly in the regime where fewer transitions at equilibrium are a safer estimate.

\renewcommand{\thefigure}{S\arabic{figure}}
\begin{figure}[H]
    \centering
    \includegraphics[width=\textwidth]{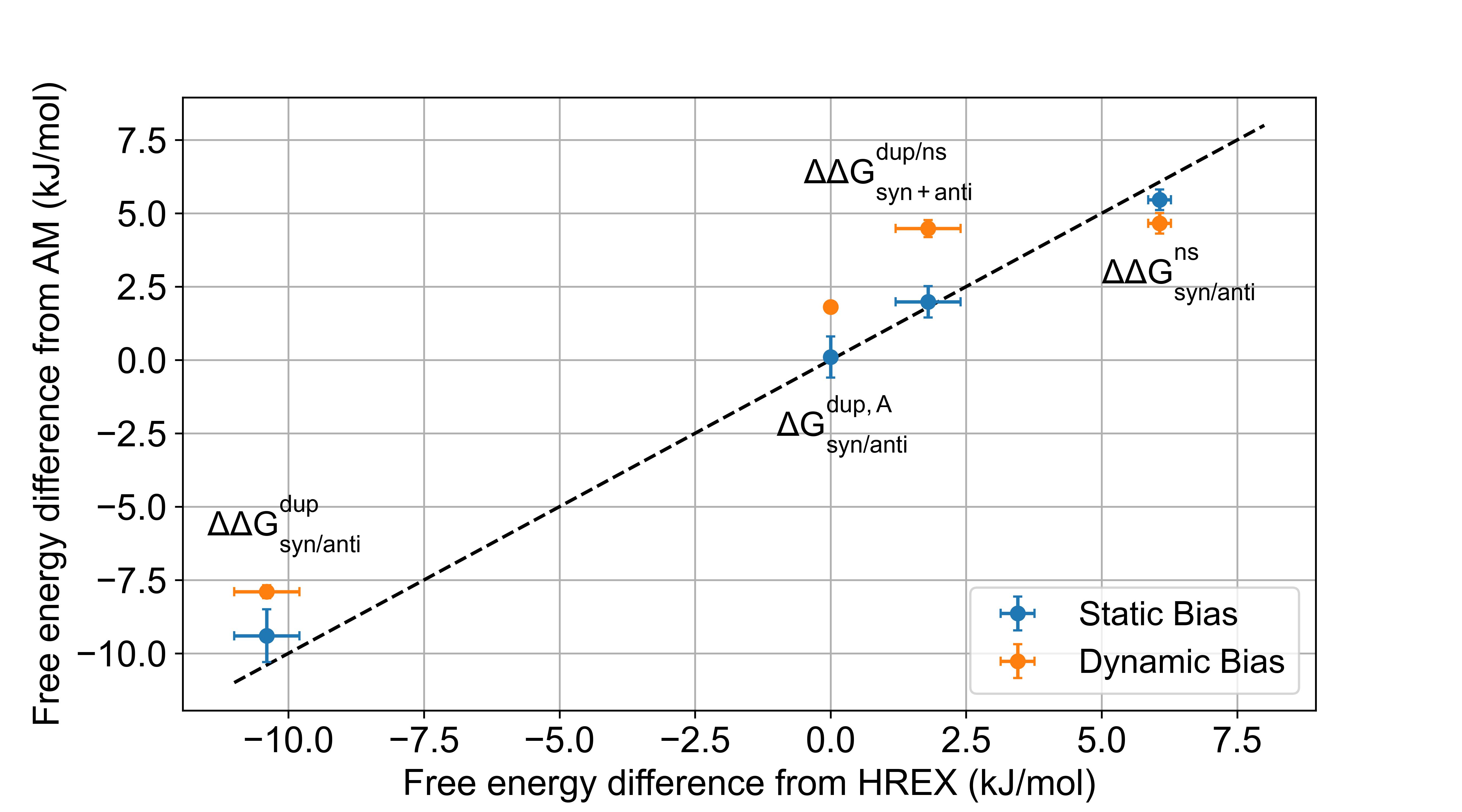}   
    \caption{Comparison of free energy differences computed in Ref.~\cite{piomponi2022molecular} with Hamiltonian replica exchange (HREX) and $\Delta \Delta G$ computed with alchemical metadynamics (AM) in this work, for two cases: (1) static bias (as discussed in the main text) and (2) dynamic bias.}
    \label{compare_ACS}
\end{figure}

\section{Supplementary Figures}
\renewcommand{\thefigure}{S\arabic{figure}}
\begin{figure}[H]
    \centering
    \includegraphics[width=\textwidth]{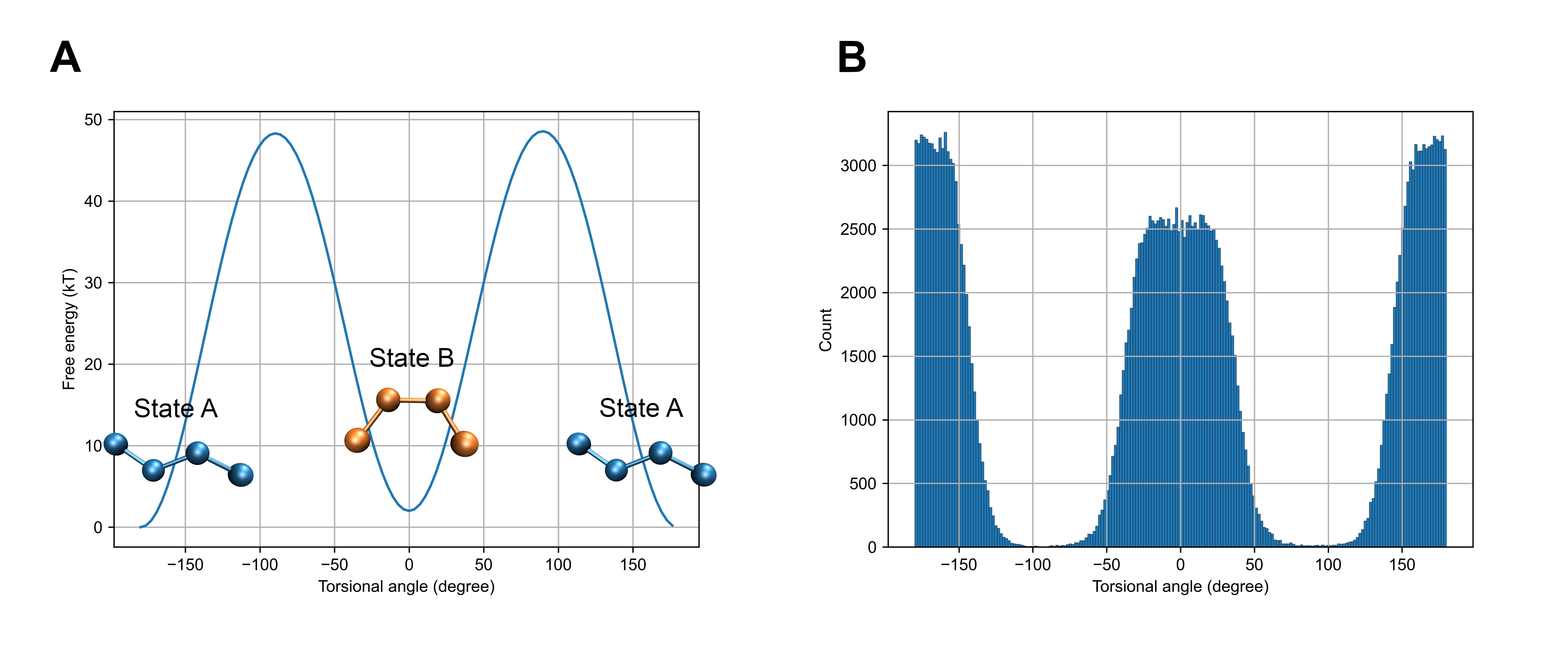}   
    \caption{(A) The free energy profile as a function of the torsional angle. We refer the structures have a torsional angle of $\pm$180$^{\circ}$ and 0$^{\circ}$ as State A (trans isomer) and State B (cis isomer). The torsional free energy barrier starting from either state is around 48.56 kT, which might not be exact since the analysis was done on a very short (5 ns) simulation solely for generating configurations at both states. (B) The histogram of the sampled torsional angle in the torsional metadynamics. As can be seen, the system was able to sample both states frequently during the short simulation.}
    \label{sys2_torsional_MetaD}
\end{figure}

\renewcommand{\thefigure}{S\arabic{figure}}
\begin{figure}[H]
    \centering
    \includegraphics[width=\textwidth]{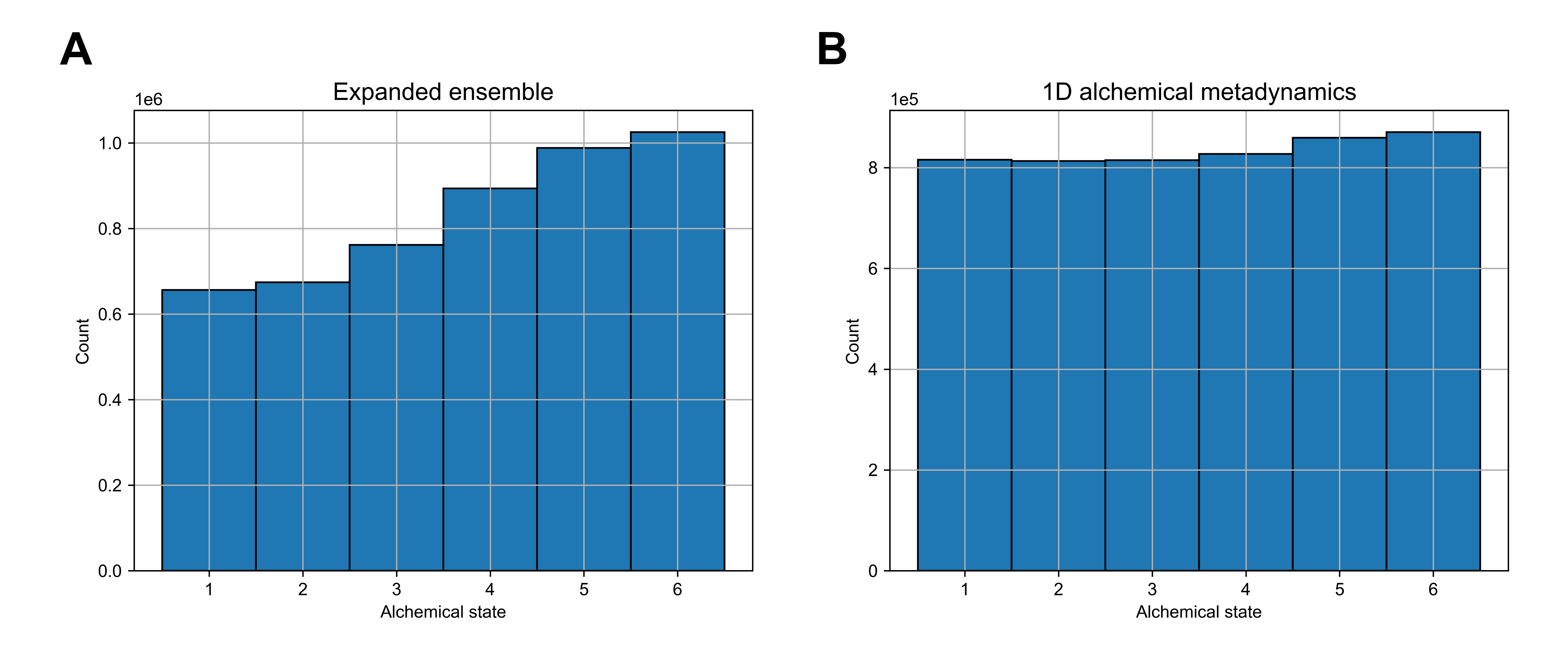}   
    \caption{The histograms of the state visitation in (A) expanded ensemble and (B) 1D alchemical metadynamics of System 1. Both simulations were able to sample all the intermediate states frequently.}
    \label{sys1_hist}
\end{figure}

\renewcommand{\thefigure}{S\arabic{figure}}
\begin{figure}[H]
    \centering
    \includegraphics[width=\textwidth]{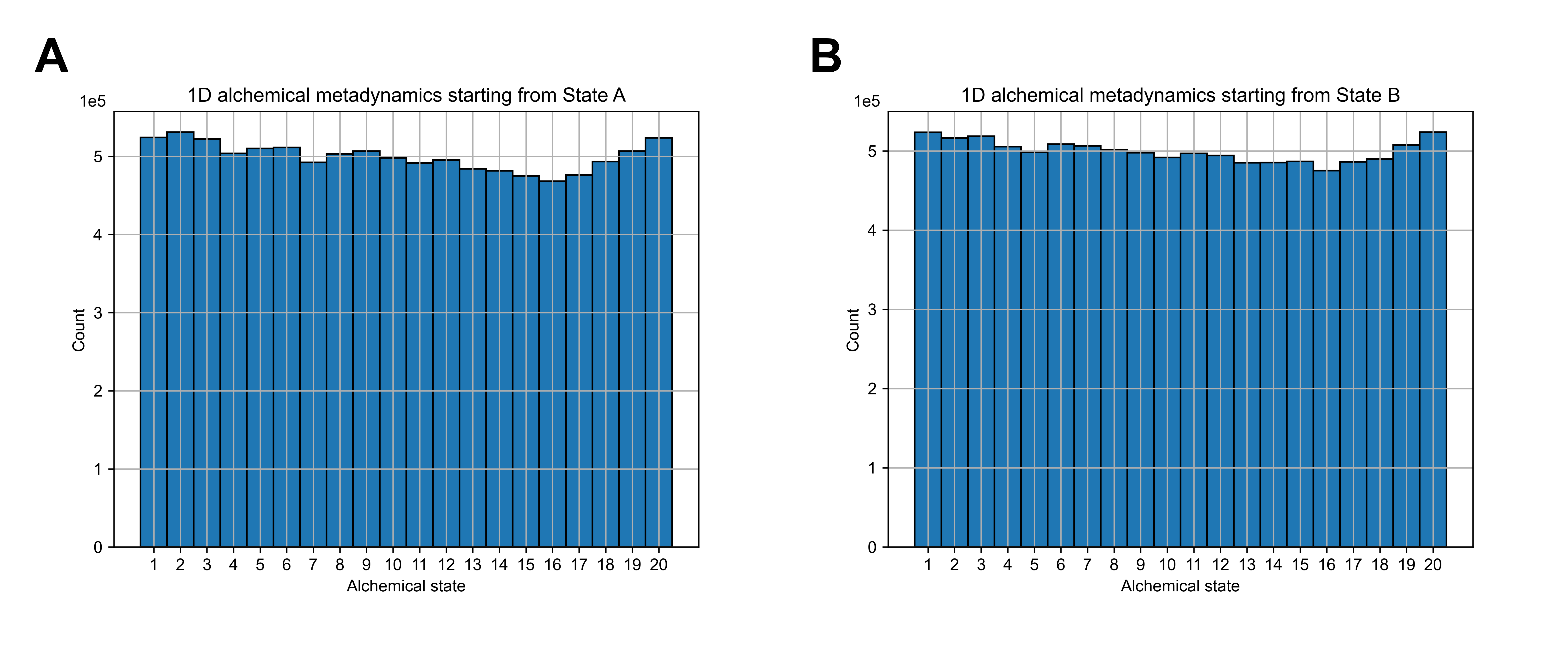}   
    \caption{The histograms of the state visitation in the 1D alchemical metadynamics starting from (A) State A and (B) State B. Both simulations were able to freely sample the alchemical space.}
    \label{sys2_1D_hist}
\end{figure}

\renewcommand{\thefigure}{S\arabic{figure}}
\begin{figure}[H]
    \centering
    \includegraphics[width=\textwidth]{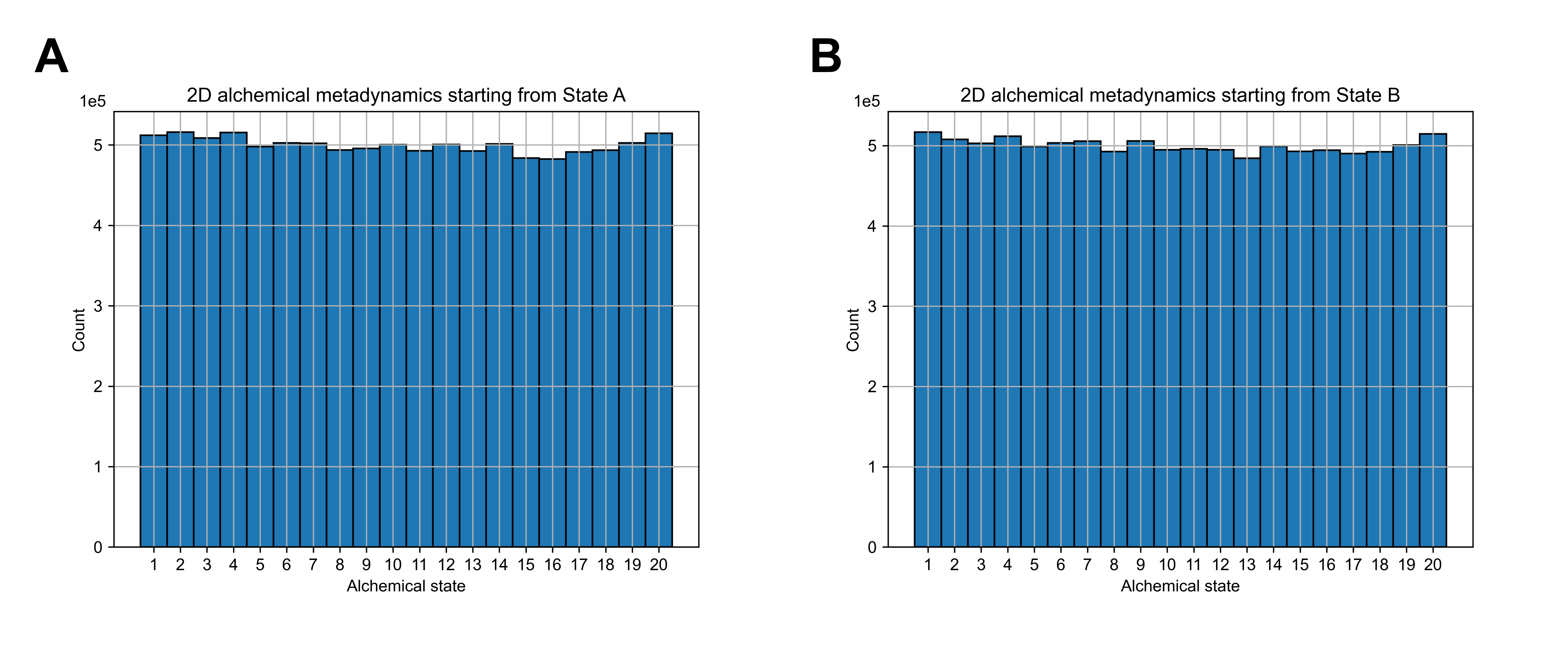}   
    \caption{The histograms of the state visitation in the 2D alchemical metadynamics starting from (A) State A and (B) State B. Similar to the two 1D simulations of System 2, both 2D simulations were able to freely sample the alchemical space.}
    \label{sys2_2D_hist}
\end{figure}

\renewcommand{\thefigure}{S\arabic{figure}}
\begin{figure}[H]
    \centering
    \includegraphics[width=\textwidth]{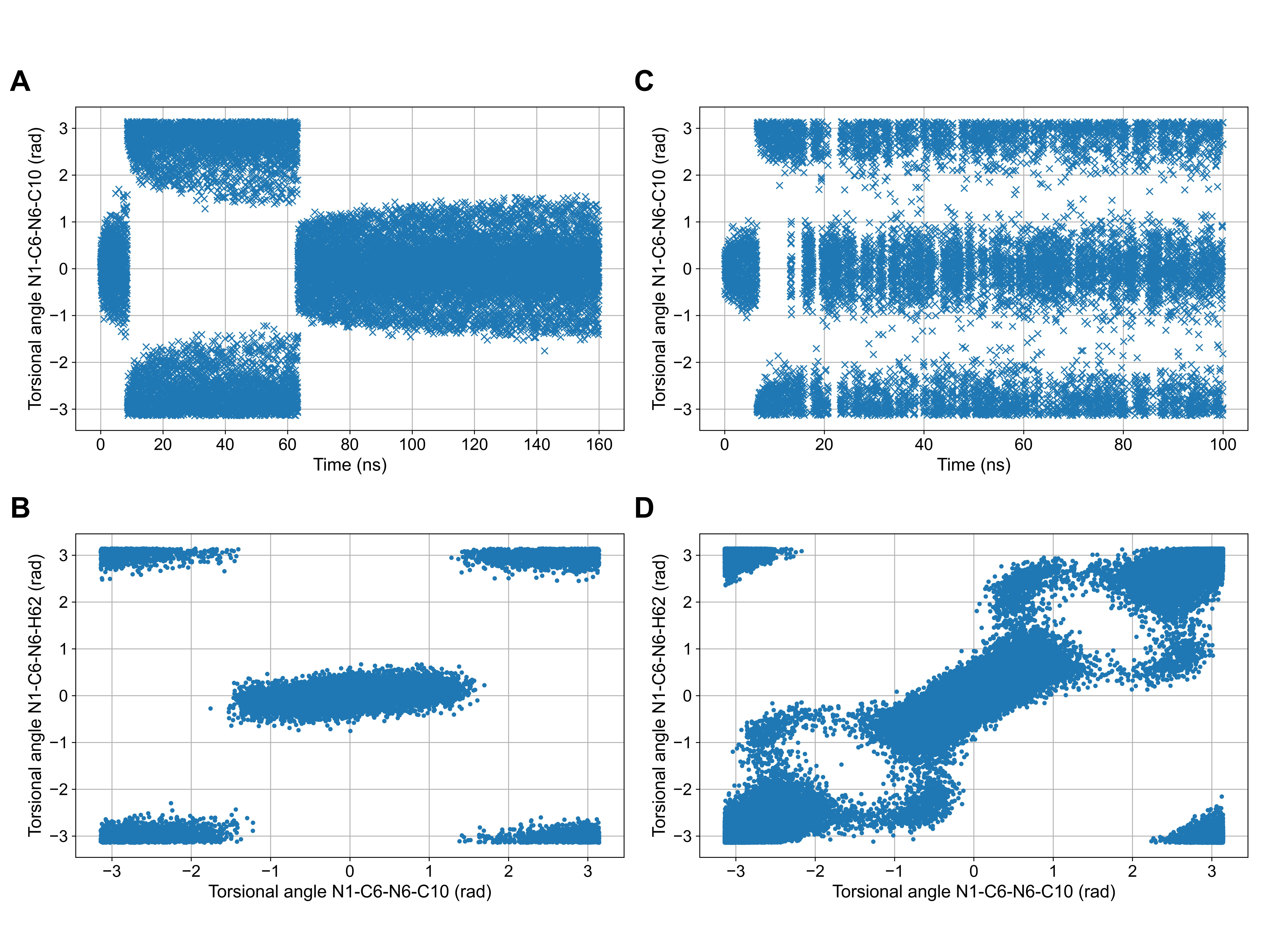}   
    \caption{(A) Value of the torsional angle N1-C6-N6-C10 as a function of time when the same torsional is used as CV, in a simulation performed at dynamic bias potential. In the 160 ns of simulations, the system only switches once from syn to anti state after about 8 ns and then back to syn after about 60 ns (B) Value of the torsional angle N1-C6-N6-C10 as a function of time when an averaged torsion between  N1-C6-N6-C10,  N1-C6-N6-H62, and  N1-C6-N6-H61 ($+ \pi$) is used as biasing collective variable. In this case, the system becomes diffusive on  N1-C6-N6-C10 after a few ns (C)  N1-C6-N6-C10 vs  N1-C6-N6-H62 when  N1-C6-N6-C10 is used as CV (D) N1-C6-N6-C10 vs  N1-C6-N6-H62 when the averaged torsion is used as CV. The three torsions mentioned here are coupled by an improper torsion that maintains the group C10, N6, H61, and H62 planar. The results shown here demonstrate that the improper torsion is not sufficiently stiff to maintain the consistency between the three torsions when enforcing the barrier crossing. As a consequence, the single N1-C6-N6-C10 torsion is not an optimal CV to allow a proper sampling of the torsional space.}
    \label{sampling}
\end{figure}

\clearpage
\input{supporting_info.bbl}

%% file: main.bbl
\providecommand{\latin}[1]{#1}
\makeatletter
\providecommand{\doi}
  {\begingroup\let\do\@makeother\dospecials
  \catcode`\{=1 \catcode`\}=2 \doi@aux}
\providecommand{\doi@aux}[1]{\endgroup\texttt{#1}}
\makeatother
\providecommand*\mcitethebibliography{\thebibliography}
\csname @ifundefined\endcsname{endmcitethebibliography}
  {\let\endmcitethebibliography\endthebibliography}{}

%% file: supporting_info.bbl
\providecommand{\latin}[1]{#1}
\makeatletter
\providecommand{\doi}
  {\begingroup\let\do\@makeother\dospecials
  \catcode`\{=1 \catcode`\}=2 \doi@aux}
\providecommand{\doi@aux}[1]{\endgroup\texttt{#1}}
\makeatother
\providecommand*\mcitethebibliography{\thebibliography}
\csname @ifundefined\endcsname{endmcitethebibliography}
  {\let\endmcitethebibliography\endthebibliography}{}